\pdfoutput=1

\documentclass[journal,onecolumn]{IEEEtran}

\ifCLASSINFOpdf
   \usepackage[pdftex]{graphicx}

\else

\fi

\usepackage{amsmath}

\usepackage{algorithmic}

\usepackage{threeparttable}
\usepackage{newclude}
\usepackage{multirow}
\usepackage{amsfonts}
\usepackage{amsthm}
\usepackage{amsmath}
\usepackage{verbatim}
\usepackage{xcolor}
\usepackage{textcomp}		
\usepackage{epstopdf}
\usepackage{textcomp}
\usepackage[caption=false]{subfig}

\usepackage[normalem]{ulem}

\usepackage{array}
\usepackage{tabularx}
\newcolumntype{P}[1]{>{\centering\arraybackslash}p{#1}}
\newcolumntype{M}[1]{>{\centering\arraybackslash}m{#1}}

\usepackage{amsmath} 

\usepackage{amsmath}
\usepackage{siunitx}

\usepackage[caption=false]{subfig}
\usepackage{amssymb}
\usepackage{amsmath}
\usepackage{commath}
\usepackage{graphicx,bm}
\usepackage{verbatim}
\newtheorem{remark}{Remark}
\usepackage{nomencl}
\makenomenclature
\usepackage{mathtools}

\usepackage{color,soul}

\usepackage{lipsum}
\usepackage{stfloats}

\usepackage{cite} 

\begin{document}
%

\title{A Feature Selection Method for High Impedance Fault Detection}

\author{Qiushi~Cui,~\IEEEmembership{Member,~IEEE,}
        Khalil~El-Arroudi,~\IEEEmembership{Member,~IEEE,}
        and Yang~Weng,~\IEEEmembership{Member,~IEEE}
\thanks{Qiushi Cui and Yang Weng are both with the School of Electrical, Computer and Energy Engineering, Arizona State University, Tempe, AZ, USA 85281. Khalil~El-Arroudi is with the General Electricity Company of Libya, Tripoli, Libya. (E-mail: \mbox{qiushi.cui@asu.edu}; \mbox{yang.weng@asu.edu}; \mbox{khalil.elarroudi@mail.mcgill.ca}.)}

\vspace{-5mm}
}

\maketitle

\begin{abstract}
High impedance fault (HIF) has been a challenging task to detect in distribution networks. On one hand, although several types of HIF models are available for HIF study, they are still not exhibiting satisfactory fault waveforms. On the other hand, utilizing historical data has been a trend recently for using machine learning methods to improve HIF detection. Nonetheless, most proposed methodologies address the HIF issue starting with investigating a limited group of features and can hardly provide a practical and implementable solution. This paper, however, proposes a systematic design of feature extraction, based on an HIF detection and classification method. For example, features are extracted according to when, how long, and what magnitude the fault events create. Complementary power expert information is also integrated into the feature pools. Subsequently, we propose a ranking procedure in the feature pool for balancing the information gain and the complexity to avoid over-fitting. 
For implementing the framework, we create an HIF detection logic from a practical perspective. Numerical methods show the proposed HIF detector has very high dependability and security performance under multiple fault scenarios comparing with other traditional methods.


\end{abstract}


\begin{IEEEkeywords}
High impedance fault, distribution network, data mining, feature selection.
\vspace{-5mm}
\end{IEEEkeywords}

%
\IEEEpeerreviewmaketitle

\nomenclature{$df$}{Frequency deviation.}
\nomenclature{$df/dt$}{Rate of change of frequency.}
\nomenclature{$pf$}{Power factor.}
\nomenclature{$dpf/dt$}{Rate of change of power factor.}
\nomenclature{$\phi$}{Power angle (angle between local voltage and current).}
\nomenclature{$\phi/dt$}{Rate of change of power angle.}
\nomenclature{$df/dP$}{Rate of change of frequency over active power.}
\nomenclature{$df/dQ$}{Rate of change of frequency over reactive power.}
\nomenclature{$dV/dP$}{Rate of change of each phase voltage over active power.}
\nomenclature{$dV/dQ$}{Rate of change of each phase voltage over reactive power.}
\nomenclature{$P$}{Active power.}
\nomenclature{$dP/dt$}{Rate of change of active power.}
\nomenclature{$Q$}{Reactive power.}
\nomenclature{$dQ/dt$}{Rate of change of reactive power.}
\nomenclature{$dV$}{Voltage deviation.}
\nomenclature{$V_{ph}$}{Phase voltage.}
\nomenclature{$V_{ll}$}{Line voltage.}
\nomenclature{$V_{abc}$}{Three-phase voltage.}
\nomenclature{$V_{012}$}{Sequence component of voltage.}
\nomenclature{$I_{abc}$}{Three-phase current.}
\nomenclature{$I_{012}$}{Sequence component of current.}
\nomenclature{$dI/dt$}{Rate of change of each phase current.}
\nomenclature{$\theta_{V_{2}}-\theta_{V_{0}}$}{Angle difference between negative and zero
sequence voltage.}
\nomenclature{$\theta_{I_{2}}-\theta_{I_{0}}$}{Angle difference between negative and zero
sequence current.}
\nomenclature{$\theta_{V_{012}}$}{Angle of sequence component of voltage.}
\nomenclature{$\theta_{I_{012}}$}{Angle of sequence component of current.}
\nomenclature{$\theta_{H_{V1\_012}}$}{Angle of sequence component of voltage at its first order harmonic.}
\nomenclature{$\theta_{H_{I1\_012}}$}{Angle of sequence component of current at its first order harmonic.}
\nomenclature{$H_{Vn}$}{Harmonic magnitude of voltage of the order $n$, $n=1,\dotsi,6$.}
\nomenclature{$H_{In}$}{Harmonic magnitude of current of the order $n$, $n=1,\dotsi,6$.}
\nomenclature{$dH_{Vn}/dt$}{Rate of change of harmonic magnitude of voltage of the order $n$, $n=1,\dotsi,6$.}
\nomenclature{$dV/dt$}{Rate of change of three-phase voltage.}
\nomenclature{$KF\_I\_cos\_Hn$}{In-phase component of the n-th order current, $n=1,\dotsi,6$.}
\nomenclature{$KF\_I\_sin\_Hn$}{In-quadrature component of the n-th order current, $n=1,\dotsi,6$.}
\nomenclature{$KF\_V\_sin\_Hn$}{In-quadrature component of the n-th order voltage, $n=1,\dotsi,6$.}
\nomenclature{$KF\_V\_cos\_Hn$}{In-phase component of the n-th order voltage, $n=1,\dotsi,6$.}
\nomenclature{$KF\_V\_DC$}{DC component of voltage estimated from KF.}
\nomenclature{$DI$}{Dependability index, the ratio of the of detected HIF events to the total number of HIF events.}
\nomenclature{$SI$}{Security index, the ratio of the of detected non-HIF events to the total number of non-HIF events.}
\printnomenclature

\section{Introduction}


\IEEEPARstart{H}{igh} impedance fault (HIF) normally exists in distribution power systems with voltages ranging from $4$ kV to $34.5$ kV. Upon the occurrence of HIF, its immediate vicinity is imposed with potential danger, which is hazardous to public safety. Unfortunately, HIFs cannot always be recorded in the fault report to relay engineers and the reported cases are therefore less than what line crews observe from the field. It was revealed in \cite{ref:tengdin1996high} that conventional protection cleared only $17.5\%$ of staged HIFs. With renewable integration into the distribution grids, the importance of HIF detection increases dramatically. Therefore, an effective HIF detection method is required to avoid false tripping and maintain the continuity of power supply. 

Specifically, an HIF is usually associated with an undowned or downed conductor. The undowned conductor scenario involves the contacts between overhead lines and tree limbs that have large impedance. Similarly, if a downed conductor falls on a poorly conductive surface such as sand, asphalt, grass, soil, and concrete, the fault current might be too low to reach the pickups of traditional ground overcurrent relays. Typical fault currents are reported ranging from $10$ to $50$ amps, with an erratic waveform \cite{ref:tengdin1996high}. 



It has been decades for researchers and engineers to seek for a universally effective solution to HIF detection. At the early stage, enhancements of conventional relays are proposed, leading to a proportional relaying algorithm \cite{ref:carr1981detection}, impedance-based method \cite{ref:choi2004new}, and PC-based fault locating and diagnosis algorithm \cite{ref:zhu1997automated}. However, these methods are ineffective in detecting HIFs with a low fault current. For this problem, harmonics patterns are utilized to capture HIF characteristics, such as magnitudes and angles of $3^{rd}$ and $5^{th}$ harmonics \cite{ref:yu1994adaptive}, even order harmonic power \cite{ref:kwon1991high}, and interharnomic currents \cite{ref:macedo2015proposition}. Besides, \cite{ref:girgis1990analysis} proposes a Kalman-filter-based method to monitor harmonics in HIF detection. This type of methods actively injects higher than fundamental frequency signals like positive/zero voltage signals \cite{ref:sagastabeitia2012low} into the grid to detect HIFs. Moreover, wavelet transform \cite{ref:elkalashy2008dwt}, genetic algorithm \cite{ref:haghifam2006development} and mathematical morphology \cite{ref:gautam2013detection} are proposed to detect HIFs. Unfortunately, most of these attempts at addressing HIF detection issues rely on simple thresholds and logic, which lacks a systematical procedure that determines the most effective features for various distribution systems and scenarios during HIFs. Therefore, it is getting necessary to introduce a systematic design for a learning framework so that information gain in high-dimensional correlation can be quantified for better HIF detections.

For learning, artificial intelligence such as expert system is proposed in the early 90's \cite{ref:kim1990parameter}. After this work,  methods using neural networks \cite{ref:kim1991learning}, decision trees \cite{ref:sheng2004decision} and fuzzy inferences \cite{ref:haghifam2006development}  are discussed in the subsequent years. In recent years, some data processing techniques including wavelet transform and mathematical morphology are gaining popularity in HIF detection. These techniques supply historical data to several machine learning algorithms (Bayes, nearest neighborhood rule, support vector machine (SVM), etc.) to differentiate fault cases \cite{ref:sedighi2005high,ref:lai2005high,ref:sarlak2013high}. 

Although the work above reveals the importance of machine learning in HIF detection, they only utilize a certain type of detection features on general HIFs. 
However,  
it is unlikely for a certain category to capture all characteristics of HIFs. Actually, various physical features from multiple types of signal processing techniques should be generated to explore the HIF pattern. 
In addition, the important step of feature selection should not be omitted before applying any learning algorithm. Otherwise, the historical data is not utilized enough for efficient learning in HIF. 



This paper contributes to use variable-importance-based feature selection method to identify an effective feature set out from a large feature pool. Specifically, we conduct a  systematic design of HIF feature pool by looking into when the fault happens, how long it lasts, and what the magnitude of the fault is. For when, we first calculate different quantities such as active power and reactive power based on voltage and current time series. Then, we use the derivative of these quantities to tell when there is a potential change due to HIF. For how long, we use discrete Fourier transform (DFT) to quantify the harmonics so those suspicious ones can be recorded for later inspection. For what magnitude, we employ Kalman Filter (KF) based harmonics coefficient estimation. Finally, power expert information is integrated into the pool, e.g., the angle difference between zero and negative sequence voltage. 
Finally, we focus on the power of feature extraction, information ranking, and detection logic, the merits of which keep unchanged under different HIF models.

In addition to the feature pool establishment, we also provide a framework for learning: feature ranking for maximizing information gain, HIF detection logic, and performance analysis. Comparing to the signal processing techniques in \cite{ref:elkalashy2008dwt,ref:haghifam2006development,ref:gautam2013detection}, the applications of DFT and KF in this paper are mature, simple, cheap and reliable, which are widely deployed in present digital relays \cite{ref:sel2013421relay} and PMUs \cite{ref:visimax2015pmu}. 

This paper is organized as follows: Section \ref{sec:HIFmodel} introduces three types of HIF models. Section \ref{sec:proposedfeatures}, \ref{sec:HIFlogic} and \ref{sec:DTLearning} elaborate on the proposed systematical method of detecting HIFs, from the feature selection method to the generation of detection logic, and the suggested performance analysis. The conclusions are presented in the last section.

\section{High Impedance Fault Modeling}
\label{sec:HIFmodel}

Although HIF phenomena are difficult to model in general, there are mainly three ways to model HIFs including both downed and undowned types for analysis. Each way provides acceptable similarity with real HIFs from its own perspective. In the following, we briefly explain each of them and the motivation behind the chosen model. 

\begin{itemize}
\item The first one is called the transient analysis of control systems (TACS) controlled switch, as proposed in \cite{ref:wai1998novel}. This model emulates arc conduction, re-ignition and extinction. The advantage of this model is the adjustable phase difference between the applied voltage and fault current. 
\item The second way originates from the Kizilcay model \cite{ref:kizilcay1991digital}， which utilizes a dynamic arc model derived from the viewpoint of control theory based on the energy balance in the arc column \cite{ref:zhang2016modelbased}.
\item The third way of modeling HIF is the employment of two anti-parallel DC-sources connected via two diodes, plus two variable resistors. The nonlinear impedances was included to add the non-linearity of fault current \cite{ref:yu1994adaptive}. Later on, the model is extended with two anti-parallel DC-sources connected via two diodes \cite{ref:emanuel1990high}, which modeled the asymmetric nature, as well as the intermediate arc extinction around current zero. The above model was then modified by adding one \cite{ref:sheng2004decision} or two \cite{ref:lai2005high} variable resistances in series with the DC sources. This kind of model is able to model the effective impedance and thus the randomness of the resulting fault current. 
\end{itemize}


In this paper, we employ the third model (see Appendix \ref{sec:Journal2017AppHIFmodel} for details) due to its easiness of implementation in Matlab Simulink for multiple simulations to realize the proposed machine learning-based method. In addition, this model is further improved here by replacing the two variable resistors with two controlled resistors. Each controlled resistor has an integrator to represent the moisture changing process in the vicinity of the point of contact of the conductor with the ground, a randomizer to introduce more randomness during HIF and a first-order transfer function to tune the response to the introduced randomness.

\begin{figure}[!htb]
	\centering
	\includegraphics[width=3.9in]{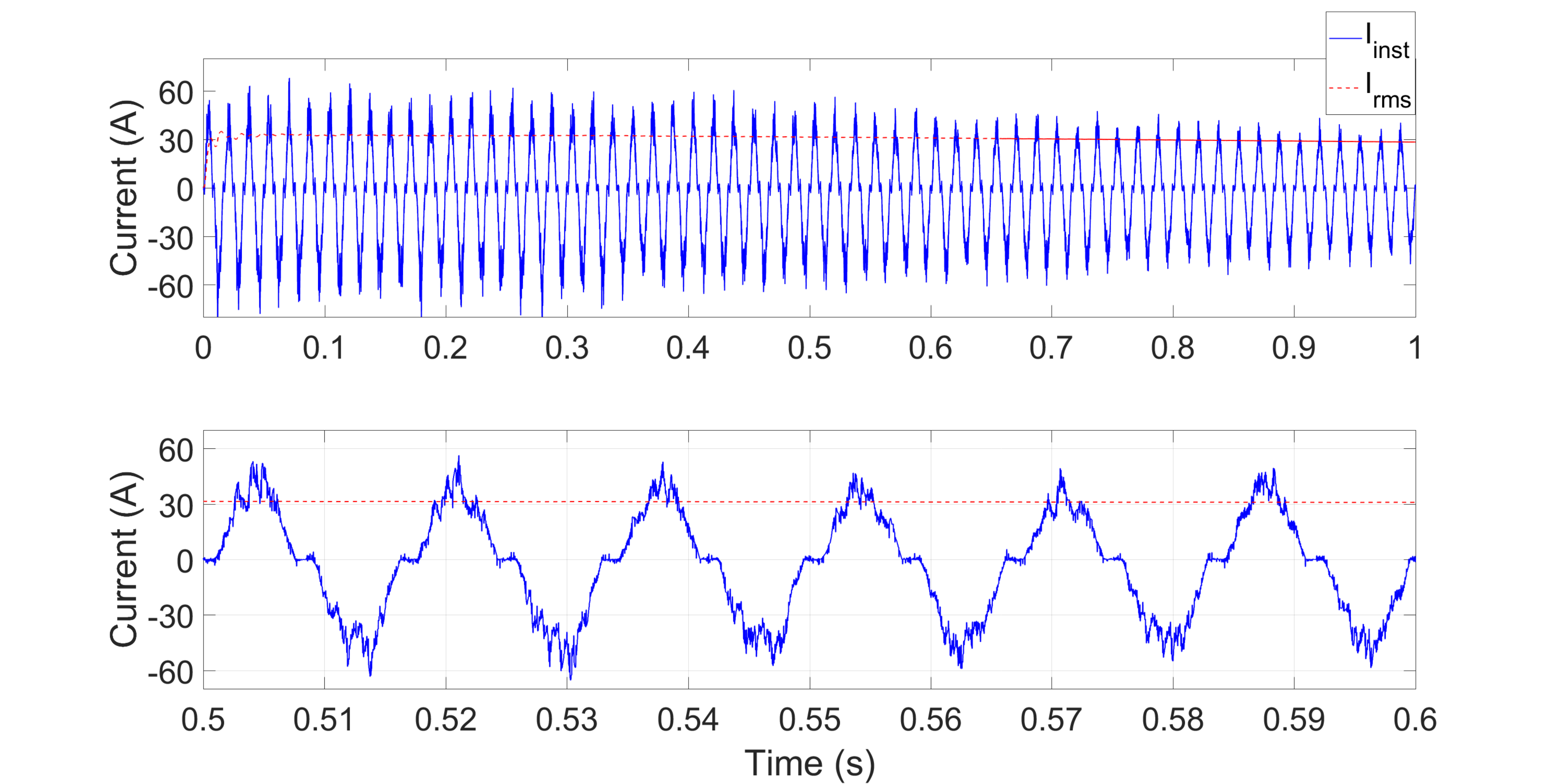}
	\centering
	\caption{The current waveforms upon HIF. The upper waveform shows the instantaneous and RMS HIF currents during $1$ sec. The lower waveform is zoomed in from the upper waveform from $0.5$ to $0.6$ sec.}
	\label{fig:hifcurrentmodel}
\end{figure}

This model therefore becomes more accurate than the one in \cite{ref:lai2005high} since the moisture change and system dynamic response are incorporated. The obtained HIF current waveforms are presented in Fig. \ref{fig:hifcurrentmodel}, which clearly displays the irregular, random, asymmetric and decreasing current waveforms upon the HIF. Meanwhile, the course of arc extinction is depicted as well around small current in this figure. It not only highlights the capability of the employed model on arc extinction modeling but also the moisture changing process. Test results of this HIF model reveal a good modeling performance and are validated in the simulation \cite{ref:gautam2013detection} and field test results \cite{ref:elkalashy2007modeling}.





\vspace{-4mm}
\section{Feature Selection Method for High Impedance Fault}
\label{sec:proposedfeatures}

Feature selection helps HIF detection identify key feature set and reduce data amount/layers, which increases the applicability of the method. Therefore, we elaborate on the way of selecting the proposed key features in this section. The variable-importance in feature evaluation is firstly explained, followed by the pool of features and selected features. The way of obtaining the feature pool data is highlighted in the end.

\vspace{-4mm}
\subsection{Variable-importance in Feature Evaluation}
\label{sec:variableimportance}

The decision-tree-based algorithm in machine learning provides protection engineers with optimal relay logic and settings in distribution network protection \cite{ref:cui2017islanding}. However, it is of significant challenge to locate the key features of HIF given its randomness and irregularity. In other words, an effective and unbiased feature evaluator is required to calculate the merit of each tested feature before the classification between HIF event and non-HIF event. Here, we take advantages of the information gain and minimum description length (MDL)-based discretization algorithm to select important features during HIF. For the convenience of power background readers, we call MDL score the variable of importance in this paper. 

The MDL-based method relies on the information gain (also known as entropy). Once the information gain of each feature is calculated for the classification variable, those features that contribute more information will have a higher information gain value over others, whereas those that do not add much information will have a lower score and can be removed. 




The score of variable-importance is one type of selection measures in machine learning. The problem of selecting the best attribute can be stated as the problem of selecting the most compressive attribute \cite{ref:ming1997introduction}. Assuming that all features are discrete, the objective is to find the best features that maximize the selection measure. ``$n..$" denotes the number of training instances and ``$n_{i.}$" is the number of training instances from class $C_i$, $n_{.j}$ is the number of instances with the $j$-th value of the given attribute, and $n_{ij}$ is the number of instances from class $C_i$ and with the $j$-th value of the given attribute. Given $C$ classes, the $MDL$ can be defined as follows using the logarithm of all possible combinations of class labels:

\begin{equation}\label{eq:paretomle2}
\begin{aligned}
\text{MDL}=\frac{1}{n..}\Big(\tbinom{n..}{n_{1.},...,n_{C.}}-\sum_{j} \log \tbinom{n_{.j}}{n_{1j},...,n_{Cj}} \\ +\log \tbinom{n..+C-1}{C-1}-\sum_{j} \log \tbinom{n_{.j}+C-1}{C-1}\Big)
\end{aligned}
\end{equation}

In this paper, we use the $MDL$ value to differentiate the merit of each detection feature for the classification between HIFs and non-HIFs. More details regarding the derivation of (\ref{eq:paretomle2}) can be found in Appendix \ref{sec:variableimportanceappendix}.



\subsection{The Pool of Candidate Features}
\label{sec:candidatafeature}

In this study, $245$ features are investigated as candidate features. For example, the feature pool in Table \ref{tab:featuretype} is designed in four steps. Firstly, the time series data of voltage and current is obtained through with the DFT-based technique. Upon the aforementioned data, the feature pool is greatly expanded with multiple physical quantities through calculation in the second step. These calculated measurements range from the basic value (e.x. $df$, frequency) to the first order derivative (e.x. $df/dt$, the rate of change of frequency), considering both the absolute value and its changing rate. Thirdly, the harmonic coefficients are estimated through the KF-based technique, presenting the in-phase and in-quadrature components, i.e. $KF\_I\_sin\_H1$. Lastly, in order to capture some unconventional phenomena, some features are invented in the category of "other feature". For example, $\theta_{V_{2}}-\theta_{V_{0}}$, the angle difference between the negative and zero sequence voltage, is a good indicator of the unbalance level in distribution grids. Note that harmonic phase angles are in harmonic degrees and are the phase difference between the zero crossing of the fundamental frequency reference and the next zero crossing in the same direction of the harmonic.

\begin{remark}
These features are extracted mainly through two techniques: discrete Fourier transform (DFT) and Kalman filter (KF). Both techniques are simple, reliable and implementable in engineering fields. The DFT is used to capture the majority of physical quantities in fault detection as is widely used in microprocessor-based relays. On the other hand, the utilization of the KF-based algorithm is motivated by the fact that it can accurately track the harmonics and inter-harmonics coefficients at given frequency components embedded in the input signals.
\end{remark}

\vspace{-5mm}

\begin{table}[!hbt]
	\renewcommand{\arraystretch}{1.3}
	\caption{Feature Pool.}
	\label{tab:featuretype}
	\centering
    
	\begin{tabular}{ccc}
		\hline
		\hline
		Feature Type & Designed Feature & Other Feature\\
				
		\hline

		\multirow{7}{*}{DFT-based} & $df, df/dt, P, dP/dt, pf, $ & $df/dP, df/dQ,$ \\
		
		& $dpf/dt, Q, dQ/dt, \phi, d\phi/dt, $ & $dV/dP, dV/dQ,$ \\
		& $H_{V1}\sim H_{V6}, H_{I1}\sim H_{I6},$ & $dH_{V1}/dt\sim dH_{V6}/dt,$\\
		& $V_{abc}, V_{012}, I_{abc}, I_{012}, dI/dt, $ & $\theta_{V_{2}}-\theta_{V_{0}},\theta_{I_{2}}-\theta_{I_{0}} $ \\
		& $dV/dt, V_{ph}, V_{ll}, \theta_{V_{012}}, \theta_{I_{012}}, $ & \\				
		& $\theta_{H_{V1\_012}}, \theta_{H_{I1\_012}}$ & \\
		
		\hline
		\multirow{4}{*}{KF-based} & $KF\_I\_cos\_H1\sim H6, $		& $KF\_V\_DC$\\

		& $KF\_I\_sin\_H1\sim H6, $ & \\
		& $KF\_V\_cos\_H1\sim H6, $ & \\
		& $KF\_V\_sin\_H1\sim H6 $ & \\
		\hline		
		\hline
	\end{tabular}
	\vspace{-3mm}
\end{table}

\vspace{-2mm}

\subsection{Systems and Events for Feature Selection}
\label{sec:syseveanalysis}

\subsubsection{Benchmark System}
\label{sec:benchmarksys}

The benchmark system utilized can be found form Appendix \ref{sec:appbenchmarksys}. The system configuration under different DER Technologies is presented in Table \ref{tab:sysconfig}. 

\begin{table}[!hbt]
	\renewcommand{\arraystretch}{1.3}
	\caption{System Configuration under Different DER Technologies. }
	\label{tab:sysconfig}
	\centering
	\begin{tabular}{ccc}
		\hline
		\hline
		System Type & Location A & Location B\\
		\hline
		Synchronous-machine-based system & SG & N/A\\
		\hline
		Inverter-based system & WF & N/A\\
		\hline
		Hybrid system & SG & WF\\
		\hline
		\hline
	\end{tabular}
\begin{tablenotes}
\small
\item SG, WF and N/A stand for the synchronous generator, wind farm and ``not available`` respectively.
\end{tablenotes}
\end{table}

\subsubsection{Events Under Study}
\label{sec:syseve}

The technique is transferable on different feeders because the event category and event type in Table \ref{tab:eventcategory} are suitable for most of distribution feeders during the training. Moreover, the event category is flexible and can be tailored for other special systems by adding or deleting some of the event categories/types. In this case study, comprehensive scenarios are considered in the event category (refer to Table \ref{tab:eventcategory}). A loading condition ranging from $30\%$ to $100\%$, in a step of $10\%$, is simulated. Furthermore, eight loading conditions and three DG technologies are examined respectively on top of the base case scenario. Therefore, the number of fault and non-fault events are calculated as follows:

\begin{itemize}

  \item Fault event: since two types of fault, summing up to $13$ cases, are included, the number of fault events with one fault impedance, one fault location and one fault impedance is $(10+3)\times8\times3=312$. Given $6$ simulated fault impedances, $4$ fault inception angles and $3$ fault locations, the total number of fault events add up to $312\times6\times4\times3=22464$.
  
  \item Non-fault event: it comprises normal state, load switching (adding and shedding) and capacitor switching events. Therefore the total number of non-fault events equals to $(1+6+2)\times8\times3=216$. 

\end{itemize}

The above event number results in an imbalanced dataset, where the number of data points belonging to the minority class (``non-fault``) is far smaller than the number of the data points belonging to the majority class (``fault``). Under this circumstance, an algorithm  get insufficient information about the minority class to make an accurate prediction. Therefore, the synthetic minority over-sampling technique (SMOTE) is employed to generate synthetic samples and shift the classifier learning bias towards minority class \cite{ref:chawla2002smote}.



\begin{table}[!hbt]
	\renewcommand{\arraystretch}{1.3}
	\caption{Event Category of System Under Study.}
	\label{tab:eventcategory}
	\centering
	\begin{tabular}{p{2.2cm}p{4.2cm}p{1.1cm}}
		\hline		
		\hline
		Event Category & Event Type & Number of Events\\

		\hline
		System Operating & Loading Condition (30\%-100\%) & 8\\
		
	Condition	& DER Tech. (SG, inverter, hybrid) & 3\\
		
		\hline
		\multirow{5}{*}{Fault Event} & Type 1: SLG, LLG, LL, LLLG & 10\\
		& Type 2: Downed conductor & 3\\
		& Fault impedance & 6\\
		& Inception Angle ($\SI{0}{\degree}$, $\SI{30}{\degree}$, $\SI{60}{\degree}$, $\SI{90}{\degree}$) & 4\\
		& Fault location & 3\\
						
		\hline
		\multirow{3}{*}{Non-fault Event} & Normal State
		& 1\\
		& Load Switching & 6\\
		& Capacitor Switching & 2\\
		\hline
		\hline
		\end{tabular}
\end{table}

\subsubsection{Spatial data extraction}
The HIF detection method should include spatial data by implementing current and voltage transformers and measurement devices at a substation and the downstream of the feeder. Knowledge extracted from these measurements is able to serve data from the spatial dimension for better detection coverage. Moreover, the proposed HIF detector installed along the distribution feeder is supplementary to the devices installed near the substation. Since the further the HIF is to the substation, the lower the signal magnitude becomes if the HIF detector is installed near the substation. The signal sensitivity and accuracy issues are therefore addressed. 

\subsection{Effective Feature Set (EFS)}
\label{sec:effectiveFS}
Finally, we propose an EFS in Table \ref{tab:effectivefeatures} after mining the collected data, applying the feature ranking algorithm and selecting the effective feature set (EFS) by considering the comprehensive performance in different distribution systems in Table \ref{tab:sysconfig}. According to the mathematical formulation and physical interpretation in Section \ref{sec:variableimportance} and \ref{sec:candidatafeature}, the reasons that enhance these features to be used for the fault detection are (1) some physical quantities are statistically more relative to the classification results than others, and (2) based on the merit of each feature, the features in Table IV contribute more information gain than others.

For example, $\theta_{V_{2}}-\theta_{V_{0}}$, the angle difference between the negative and zero sequence voltage, is selected since it captures the incremental of the unbalance level contributed from HIFs to distribution grids. The using of the angle difference between zero and negative sequence voltage is inspired by the work in \cite{ref:yu1994adaptive},\cite{ref:huang2014novel}, and the practical engineering experience of the authors. To the best of our knowledge, this feature is utilized in some other fault detection application such as \cite{ref:huang2014novel}, but not in HIF detection before. Table \ref{tab:EFSreference} shows the reference to the unbalanced fault detection features in EFS.

\begin{table}[!hbt]
	\renewcommand{\arraystretch}{1.3}
	\caption{Effective Feature Set of HIF Detection in Three Types of Distribution
Systems.}
	\label{tab:effectivefeatures}
	\centering
	\begin{tabular}{cc}
		\hline
		\hline
		Fault Type & Proposed Feature \\
		
		\hline
		
		\multirow{2}{*}{SLG, LL, LLG} & $V_2, I_2, \theta_{V_{2}}-\theta_{V_{0}},\theta_{I_{2}}-\theta_{I_{0}}$ \\
		
		& $KF\_V\_cos\_H3, KF\_V\_sin\_H3$ \\
		
		\hline
		\multirow{2}{*}{LLLG} & $V_{ll}, V_{ph}, H_{V1}, \theta_{H_{V1\_1}}$\\
		
		& $KF\_I\_cos\_H1, KF\_I\_sin\_H1$ \\

		\hline		
		\hline
	\end{tabular}
\end{table}

\begin{table}[hbt]
	\renewcommand{\arraystretch}{1.2}
	\caption{Reference to the Unbalanced Fault Detection Features in EFS.}
	\label{tab:EFSreference}
	\centering
	\begin{tabular}{cc}
		\hline		
		\hline
		Feature in EFS & Reference\\
		\hline
		$V_2$ & \multirow{2}{*}{\cite{ref:yu1994adaptive}, \cite{ref:Laaksonen2017method}} \\
		$I_2$ & \\
		\hline
		$\theta_{V_2}-\theta_{V_0}$ & \multirow{2}{*}{\cite{ref:huang2014novel}, \cite{ref:yu1994adaptive}}\\
		$\theta_{I_2}-\theta_{I_0}$ &  \\
		\hline
		$KF\_V_a\_cos\_H3$ & \cite{ref:yu1994adaptive} (3rd haromnic), \cite{ref:girgis1990analysis} (KF and low-order odd\\
		$KF\_V_a\_sin\_H3$ &   harmonic), \cite{ref:kamwa20123compliance} (KF harmonic decomposition)\\
		\hline
		\hline		
	\end{tabular}
\end{table}


\section{High Impedance Fault Detection Logic}
\label{sec:HIFlogic}


Inspired by the tree structure of the machine learning classifier model, the authors further explore the possibility of relating the EFS and the detection logic using simple thresholds as most of the commercial products \cite{ref:sel2016arc} and patents do \cite{ref:russell1984high}. Statistically, since three-phase faults take up only $2\%-3\%$ of the fault occurrences \cite{ref:blackburn2006protective}, an HIF detection logic is designed in this regard for unbalanced HIF only. 



\begin{figure}[!htb]
	\centering
	\includegraphics[width=3.6in]{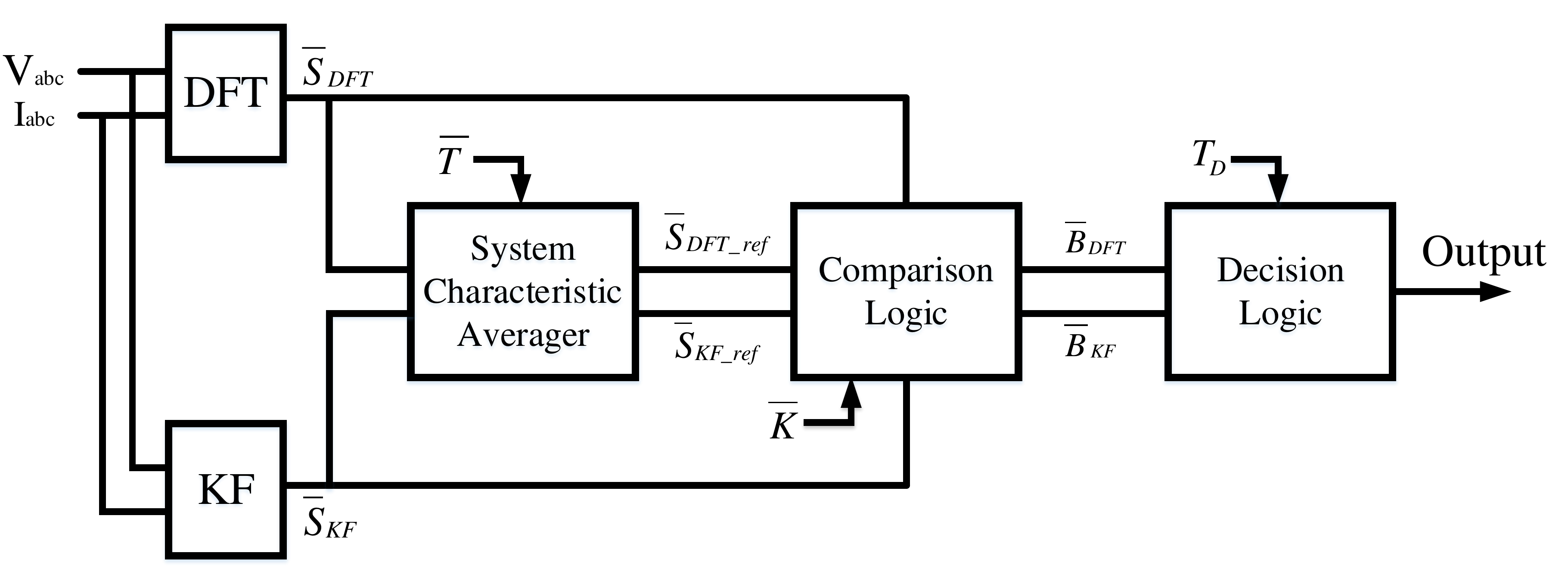}
	\caption{The proposed HIF detection logic scheme.}
	\label{fig:HIFlogic_general}
	\vspace{-3mm}
\end{figure}

The HIF detection logic is targeted to be implemented in a microprocessor-based digital relay, as guaranteed by the selected feature selection techniques discussed in Section \ref{sec:candidatafeature}. Similar to conventional digital relays, the proposed relay logic takes the voltage and current signals as its input. In addition, DFT and KF are required for corresponding feature extraction. Before the explanation of the HIF detection logic, the logic circuit is presented first in Fig. \ref{fig:HIFlogic_general}. Generally, the proposed HIF detection scheme updates its comparison and decision logic according to the obtained decision tree structure.


As indicated in the detection logic, three-phase voltage and current signals are sent to DFT and KF for feature extraction. This section takes the obtained EFS in Section \ref{sec:proposedfeatures} as an example. (\ref{eq:sdft}) and (\ref{eq:skf}) show the extracted instantaneous signals after the DFT and KF blocks:
\vspace{-2mm}
\begin{equation}
\bar{S}_{DFT}=\{s_1,s_2\,s_3,s_4\}=\{V_2,I_2,\theta_{V_2}-\theta_{V_0},\theta_{I_2}-\theta_{I_0}\}
\label{eq:sdft}
\vspace{-1mm}
\end{equation}

\begin{equation}
\vspace{-2mm}
\begin{aligned}
\bar{S}_{KF} & =\{s_5,s_6,s_7,s_8,s_9,s_{10}\}= \{KF\_V_a\_cos\_H3, \\
      & KF\_V_b\_cos\_H3, KF\_V_c\_cos\_H3,KF\_V_a\_sin\_H3, \\
      & KF\_V_b\_sin\_H3,KF\_V_c\_sin\_H3\}
\end{aligned}
\label{eq:skf}
\end{equation}

\vspace{-5mm}

\subsection{System Characteristic Averager}
The System Characteristic Averager has a memory that stores the signals for a predefined duration of $\bar{T}=\{t_1,t_2,t_3,t_4,t_5,t_6$,
$t_7,t_8,t_9,t_{10}\}$. In other words, $\bar{T}$ is the time constant that is a vector of ten elements associated with $\bar{S}_{DFT}$ and $\bar{S}_{KF}$. The signal is firstly calculated and stored at 5 minutes interval \cite{ref:yu1994adaptive}. Simulation or experimental results are then provided to validate the effectiveness of this time constant over a large time scale. In the end, each time constant is either increased or decreased depending on the signal's slow or fast dynamic process.


To avoid signal spikes, a limiter is implemented for each signal channel. Meanwhile, the time constant $\bar{T}$ is set according to the system characteristics of each individual signal. A small $t_i$ ($i=1,2,\dotsi,10$) can avoid severe step change of signal but a large $t_i$ costs more data storage and computational efforts.  The output of the System Characteristic Averager block provides the reference value $s_{i\_ref}$ ($i=1,2,\dotsi,10$) for the Comparison Logic. A reliable average value is a prerequisite to successful detection. 

\subsection{Comparison Logic}

\begin{figure}[!htb]
	\centering
	\includegraphics[width=2.4in]{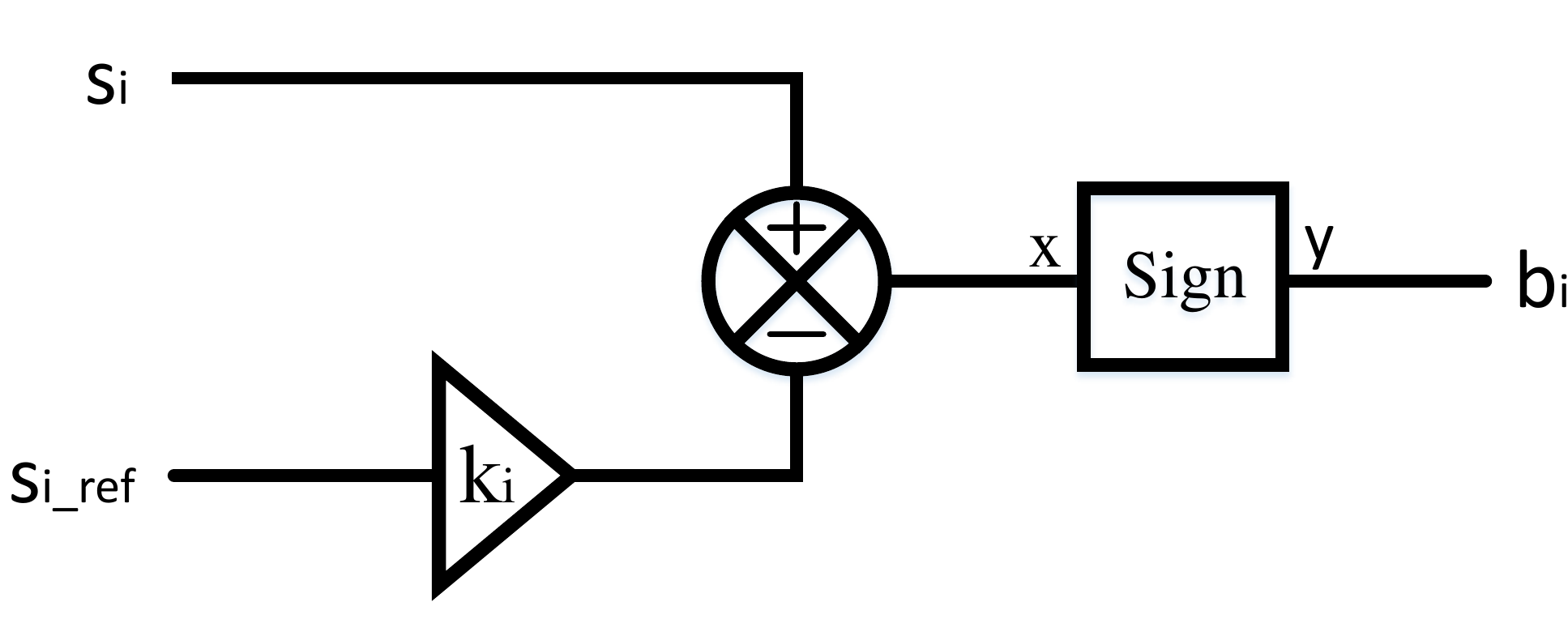}
	\caption{Comparison Logic in the proposed HIF detection logic.}
	\label{fig:comp_logic}
\end{figure}

The block of Comparison Logic is depicted in Fig. \ref{fig:comp_logic}. Based on the feature extraction technique discussed in Section \ref{sec:proposedfeatures}, the extracted instantaneous signal $s_i$ can be understood as the system background signal superimposed by the extra signal contributed from the HIF behavior. The comparison is therefore made between the extracted instantaneous signal $s_i$ and its reference value $s_{i\_ref}$ \cite{ref:yu1994adaptive}. 

The sensitivity gain of $k_i$ is incorporated in order to 1) set the margin of detection and 2) add a handle to the detection sensitivity. Where the undefined parameter of $\bar{K}$ stands for:

\begin{equation}
\bar{K}=\{k_1,k_2,k_3,k_4,k_5,k_6,k_7,k_8,k_9,k_{10}\}
\end{equation}

The sensitivity gain $\bar{K}$ is set at $1.2$ (adjustable for each element). The 20\% above and below margin is adjustable and is taken as typical blackout region where the HIF tripping is not required \cite{ref:yu1994adaptive}. This $k_i$ value can be set to close to $1.0$ after getting more confidence in HIF fault detection scheme. After the summation block in Fig. \ref{fig:comp_logic}, a Sign function is employed to provide the following decision making:

\begin{itemize}
\item When $x>0$, $y=1$;
\item When $x \leqslant 0$, $y=0$. 
\end{itemize}

The output of the comparison logic is the comparison assertion bit of $b_i$ ($i=1,2,\dotsi,10$), the $\bar{B}$, which is the input to the decision logic.

\subsection{Decision Logic}

As mentioned in the previous subsection, the comparison assertion bit of $b_i$ ($i=1,2,\dotsi,10$) is the output of the comparison logic in Fig. \ref{fig:comp_logic}. The decision logic in Fig. \ref{fig:decisionlogicHIF}, is the execution part of the HIF detection logic. There are four groups of signal bits:

\begin{figure}[!htb]
	\centering
	\includegraphics[width=2.7in]{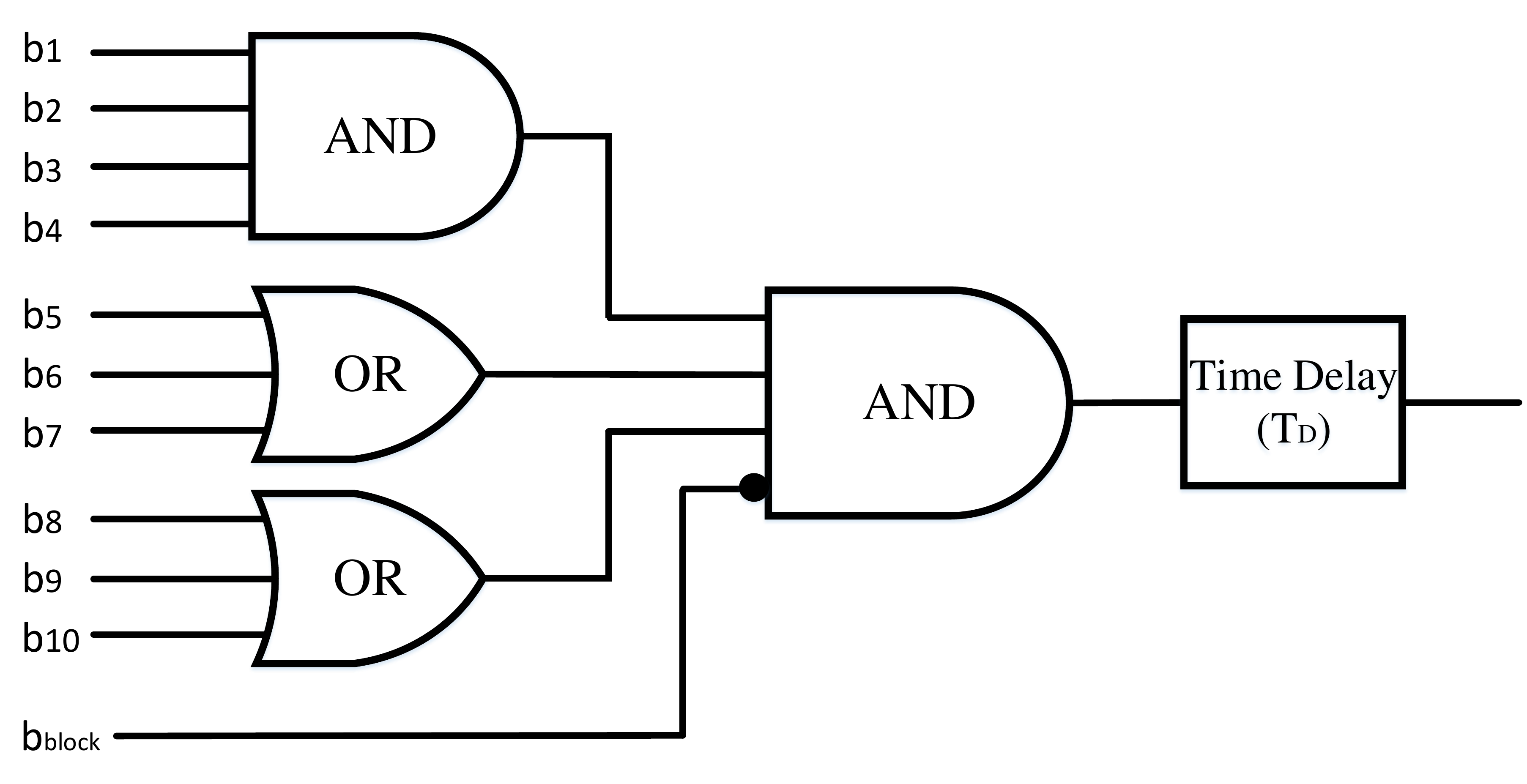}
	\caption{Decision Logic in the proposed HIF detection logic.}
	\label{fig:decisionlogicHIF}
	\vspace{-2mm}
\end{figure}

\begin{enumerate}
\item DFT-based assertion bits. The four bits go through an AND gate. If any of the four signals are not asserted, the decision logic will not be set high. 
\item KF-estimated in-phase components of third harmonic voltage. If none of the three-phase in-phase components of third harmonic estimated from the KF gets asserted, the decision logic will not be set high. 
\item KF-estimated in-quadrature components of harmonic voltage. If none of the three-phase in-quadrature components of third harmonic estimated from the KF gets asserted, the decision logic will not be set high. 
\item The blocking bit $b_{block}$. If this bit is $1$, the detection logic is blocked and none of HIF events can be detected; if this bit is $0$, HIF detection is enabled. 
\end{enumerate}

A time delay of $T_D$ is implemented because an appropriate selection of $T_D$ can effectively avoid the false operation resulting from normal switching, which sometimes contributes to third harmonics. The output of the HIF logic is either alarming or tripping signal. 

\vspace{-2mm}

\subsection{Performance Test of the Proposed HIF Detection Logic}
\label{sec:performanceHIFlogic}

The proposed HIF detection logic is tested under $7884$ new scenarios: $7776$ unbalanced faults and $108$ non-faults. The fault locations under testing include faults near B-$3$, B-$11$, and B-$19$. The detailed analysis regarding fault locations can be found in Section \ref{sec:faultlocationHIFtest}. Similar to the work in \cite{ref:yu1994adaptive} and \cite{ref:gautam2013detection}, the measurement point is at the substation. Its sampling frequency is $2000$ Hz. The time delay in Fig. \ref{fig:decisionlogicHIF} is set to $100$ ms. The average fault detection time is $0.126$ sec using OPAL-RT real-time simulator. The signals measured are the three-phase voltage and current. The features used are derived from the measured signals and can be found in the EFS in the unbalanced fault row of Table IV.

Performance comparison with the HIF detection logic in \cite{ref:yu1994adaptive}, as well as the combined conventional relay elements (frequency, over/under voltage, over current) is shown in Table \ref{tab:HIFperformancecomp}.

\begin{table}[!hbt]
	\renewcommand{\arraystretch}{1.3}
	\caption{HIF Detection Logic Performance comparison.}
	\label{tab:HIFperformancecomp}
	\centering
	\begin{tabular}{ccc}
		\hline
		\hline
		Solution under test & DI (\%) & SI (\%) \\
		\hline
		The proposed EFS and HIF detection logic &98.3&95.7 \\		
		\hline
		The HIF detection logic in \cite{ref:yu1994adaptive}& 69.0&90.7 \\	
        \hline
		Combined conventional relay elements & \multirow{2}{*}{0} & \multirow{2}{*}{98.2} \\	
        (frequency, over/under voltage, over current)& &  \\	
		\hline		
		\hline
	\end{tabular}
\end{table}


\section{Performance Analysis}
\label{sec:DTLearning}

Performance analysis includes the most commonly occurring single-line-to-ground-fault, the fault scenario analysis, and the testing results.

\vspace{-2mm}
\subsection{Single-line-to-ground Fault Analysis}



\begin{figure}[!htb]
	\centering
	\includegraphics[width=3in]{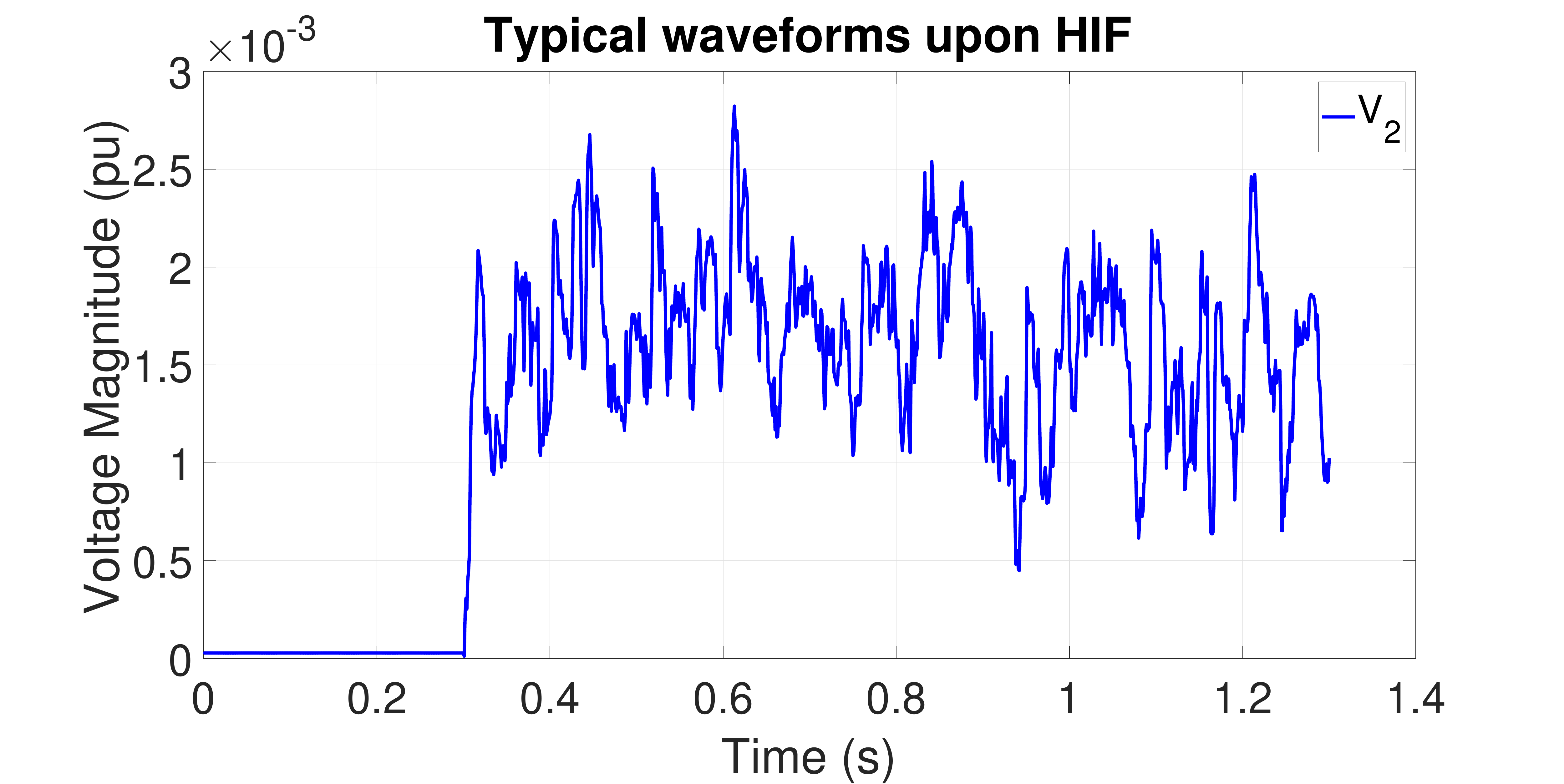}
	\caption{Typical negative sequence voltage waveforms under HIF.}
	\label{fig:v20}
\end{figure}

\begin{figure}[!htb]
	\centering
	\includegraphics[width=3in]{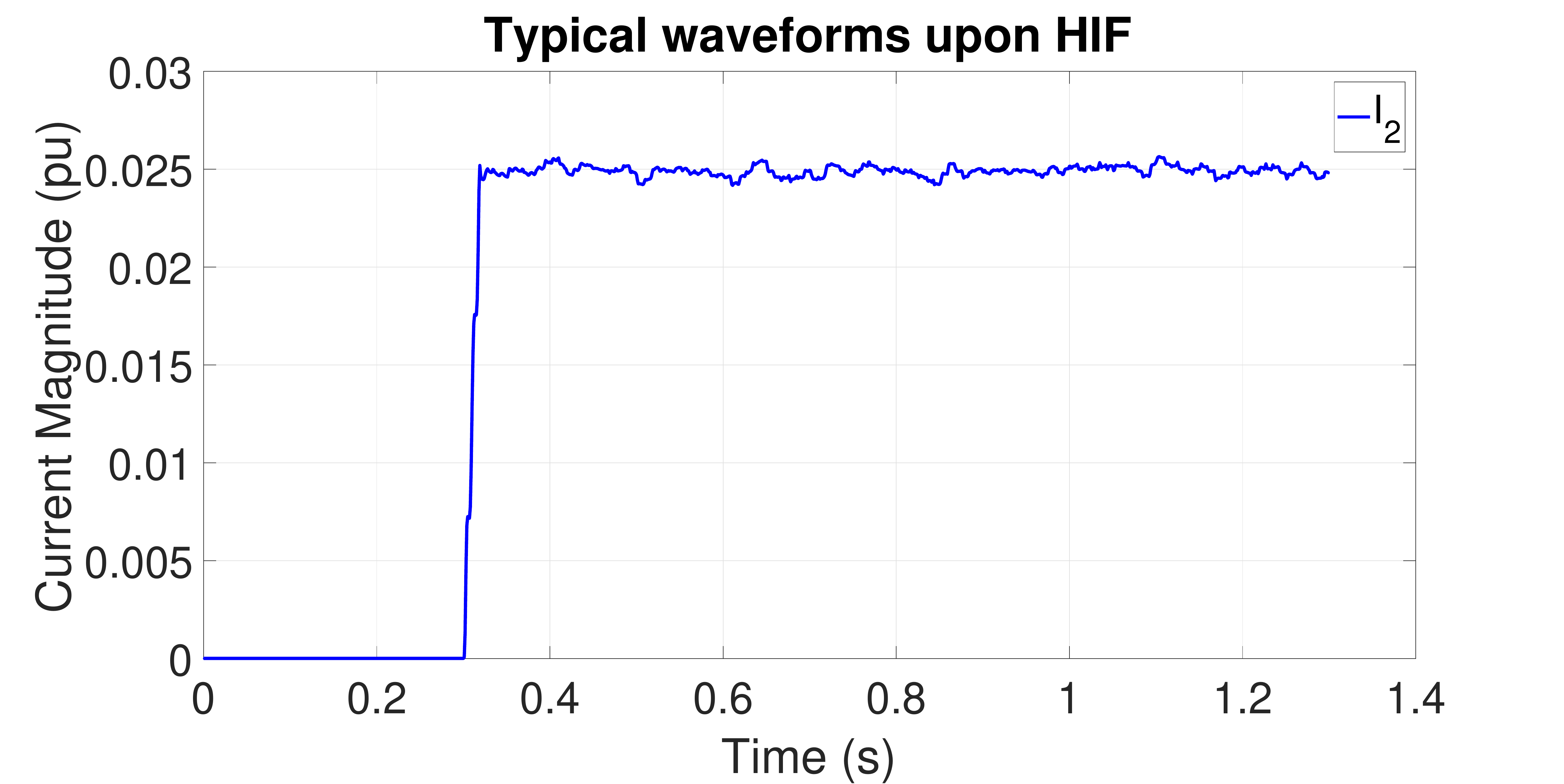}
	\caption{Typical negative sequence current waveforms under HIF.}
	\label{fig:i20}
\end{figure}

\begin{figure}[!htb]
	\centering
	\includegraphics[width=3.2in]{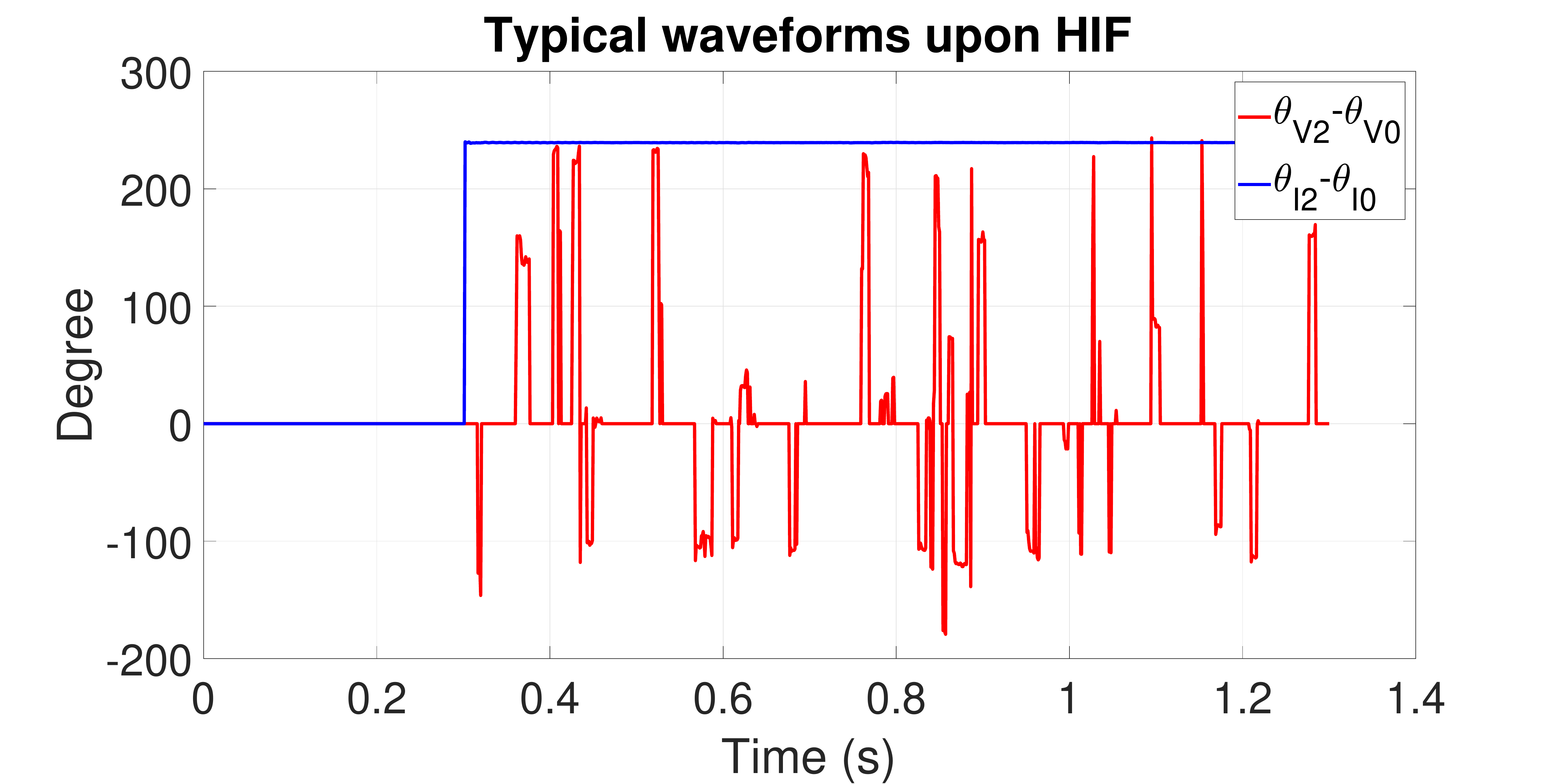}
	\caption{Typical waveforms angle difference  between zero and negative sequence voltage and current under HIF.}
	\label{fig:anglediff}
\end{figure}

\begin{figure}[!htb]
	\centering
	\includegraphics[width=3in]{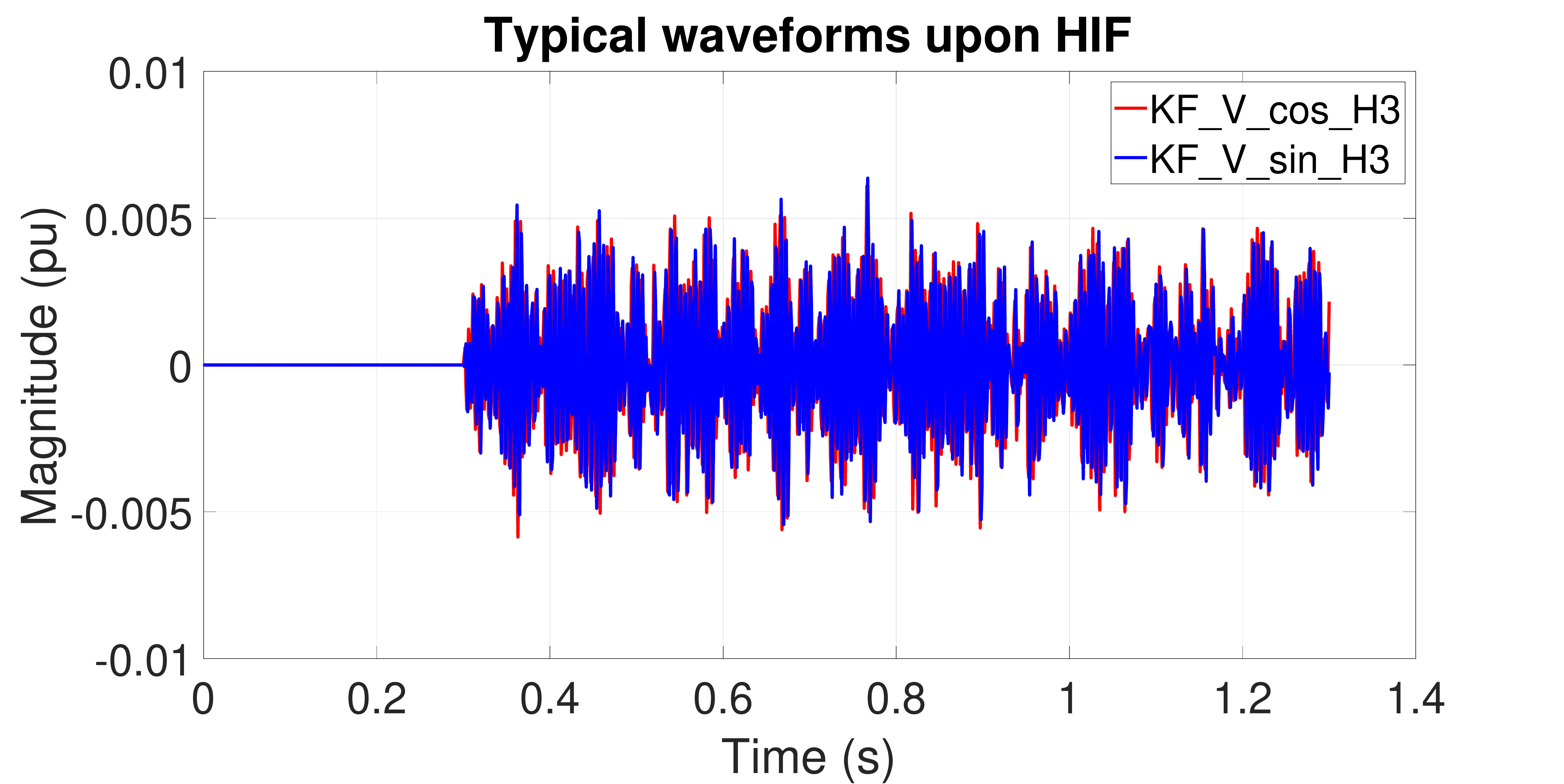}
	\caption{Typical waveforms of 3rd order in-phase and in-quadrature components under HIF.}
	\label{fig:kfh3}
\end{figure}

Typical waveforms of the proposed EFS upon single-line-to-ground fault are shown in this subsection. A single-line-to-ground HIF is applied in a hybrid distributed generation system (refer to Fig. \ref{fig:feeder_HIF}) when $t=0.3$ second. Fig. \ref{fig:v20} to \ref{fig:kfh3} shows the EFS' waveforms. 

\subsection{Fault Scenario Analysis}

We evaluate the Effective Feature Set (EFS) in terms of different fault impedances, fault inception angles, and fault locations. The quantifier for evaluation is the variable of importance explained in Section \ref{sec:variableimportance}.

\subsubsection{Fault Impedance}
\label{sec:faultimpedance}

To be practical, this paper investigates the fault impedance up to \SI{500}{\Omega} to cover typical HIFs whose fault currents are as low as $10$ amps. The variable-importance performances of each feature in EFS upon single-line-to-ground (SLG) fault, line-to-line (LL) fault, line-to-line-to-ground (LLG) fault and three-line-to-ground (LLLG) fault are all depicted in Fig. \ref{fig:Rf_slgllg}. It is concluded that:

 \begin{figure*}[htb!]
     \centering
     \subfloat[SLG fault.]{\label{fig:rfslg}\includegraphics[width=3.0in]{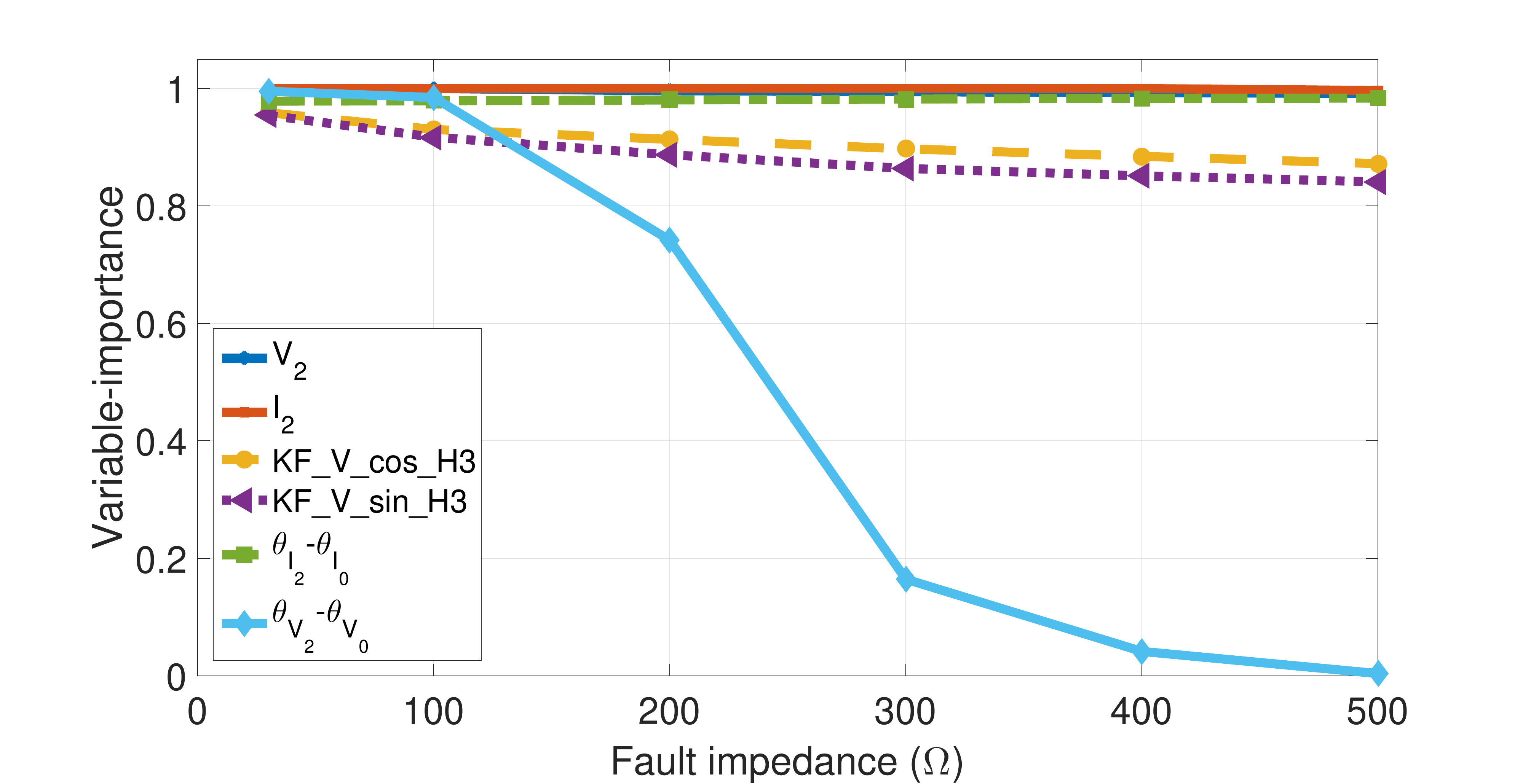}} 
     \subfloat[LL fault.]{\label{fig:rfll}\includegraphics[width=3.0in]{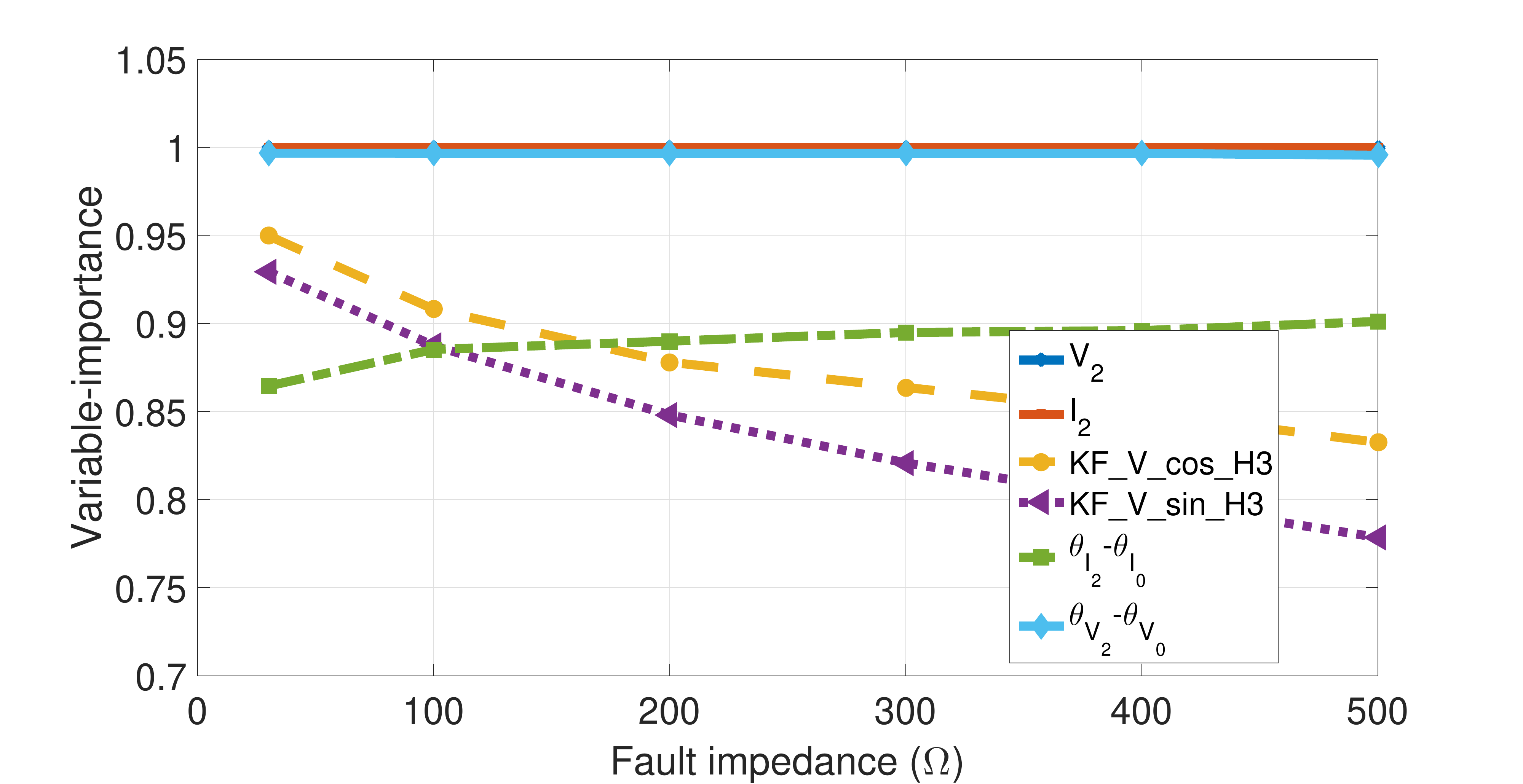}} \\
     \subfloat[LLG fault.]{\label{fig:rfllg}\includegraphics[width=3.0in]{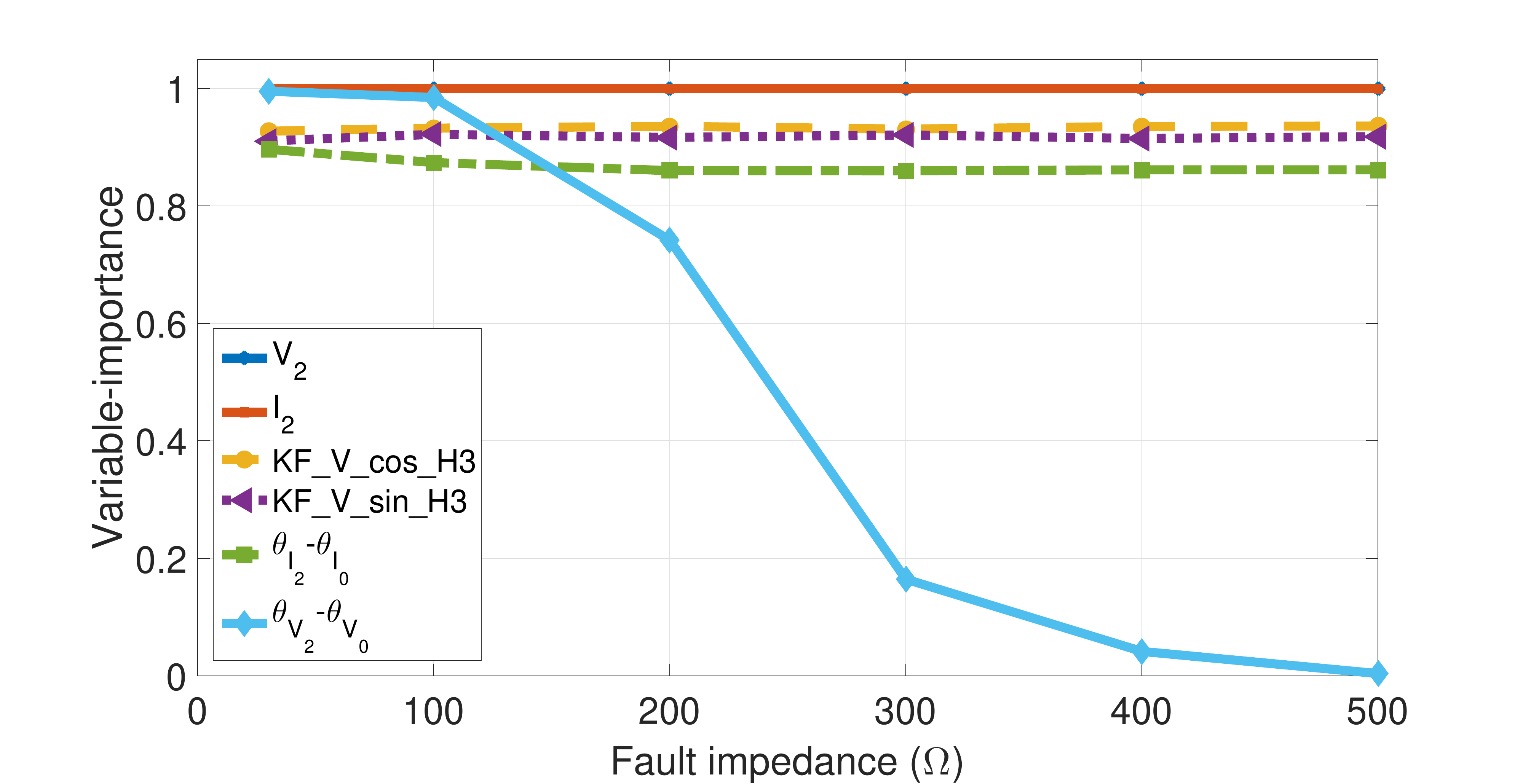}} 
     \subfloat[LLLG fault.]{\label{fig:rflllg}\includegraphics[width=3.0in]{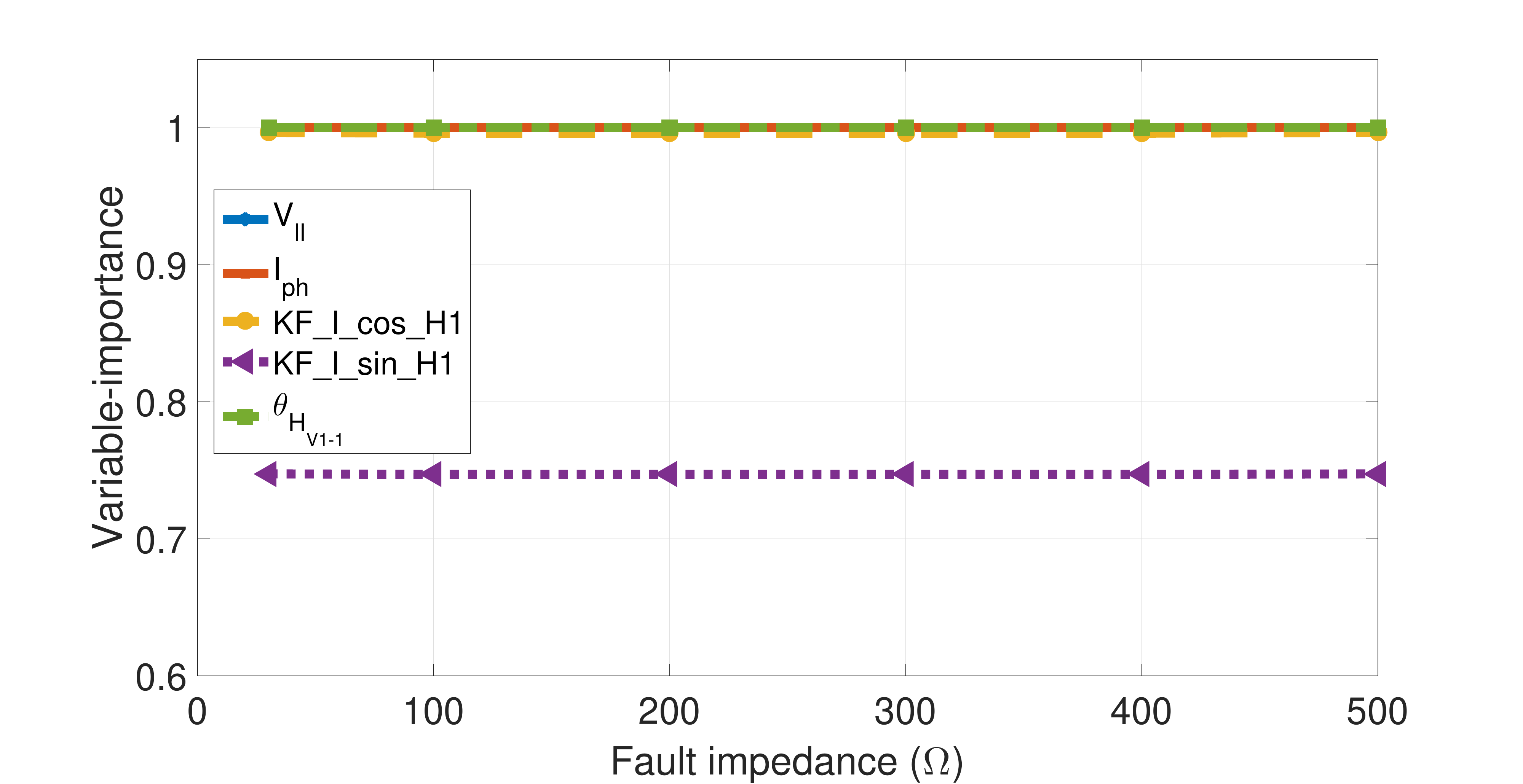}}
       \caption{Variable-importance of all features under faults in a grounded system.}
     \label{fig:Rf_slgllg}
    \vspace{-4mm}
 \end{figure*}

\begin{itemize}

  \item The negative sequence of voltage and current are most reliable features that can keep unaffected during any unbalanced fault upon a varying fault impedance (Fig. \ref{fig:Rf_slgllg}(a), to be noticed that the dark blue line for $V_2$ is covered by other lines with the value of $1$). 
  \item The feature of the angle difference between negative sequence voltage and zero sequence voltage is reliable under LL faults but vulnerable to high fault impedance under SLG and LLG faults (Fig. \ref{fig:Rf_slgllg}(a)-(d)). 
  \item The third harmonic components estimated from KF gets deteriorated when the fault impedance increases under SLG and LL faults (Fig. \ref{fig:Rf_slgllg}(a) and (b)). 
  \item The proposed three-phase HIF detection features are all performing very well except for the fundamental in-quadrature component of current estimated from KF under LLLG faults (Fig. \ref{fig:Rf_slgllg} (d)).

\end{itemize}

Furthermore, the proposed algorithm is applicable to unbalanced power systems. Since the employed feature selection method is based on the information gain, what is captured by the information gain is the incremental or variation of the negative sequence signal. Only when the variation pattern of the negative sequence feature contributes to the information gain given the output label belongs to the HIF, does this feature get selected by the proposed algorithm. As a result, the proposed method is applicable to an already unbalanced system.

\subsubsection{Fault Inception Angle}
\label{sec:faultinceptionangle}

\begin{figure*}[htb!]
     \centering
     \subfloat[Feature $\theta_{I_{2}}-\theta_{I_{0}}$ at different fault inception angles.]{\label{fig:inceptionangle_theta_i2_i0}\includegraphics[width=3.0in]{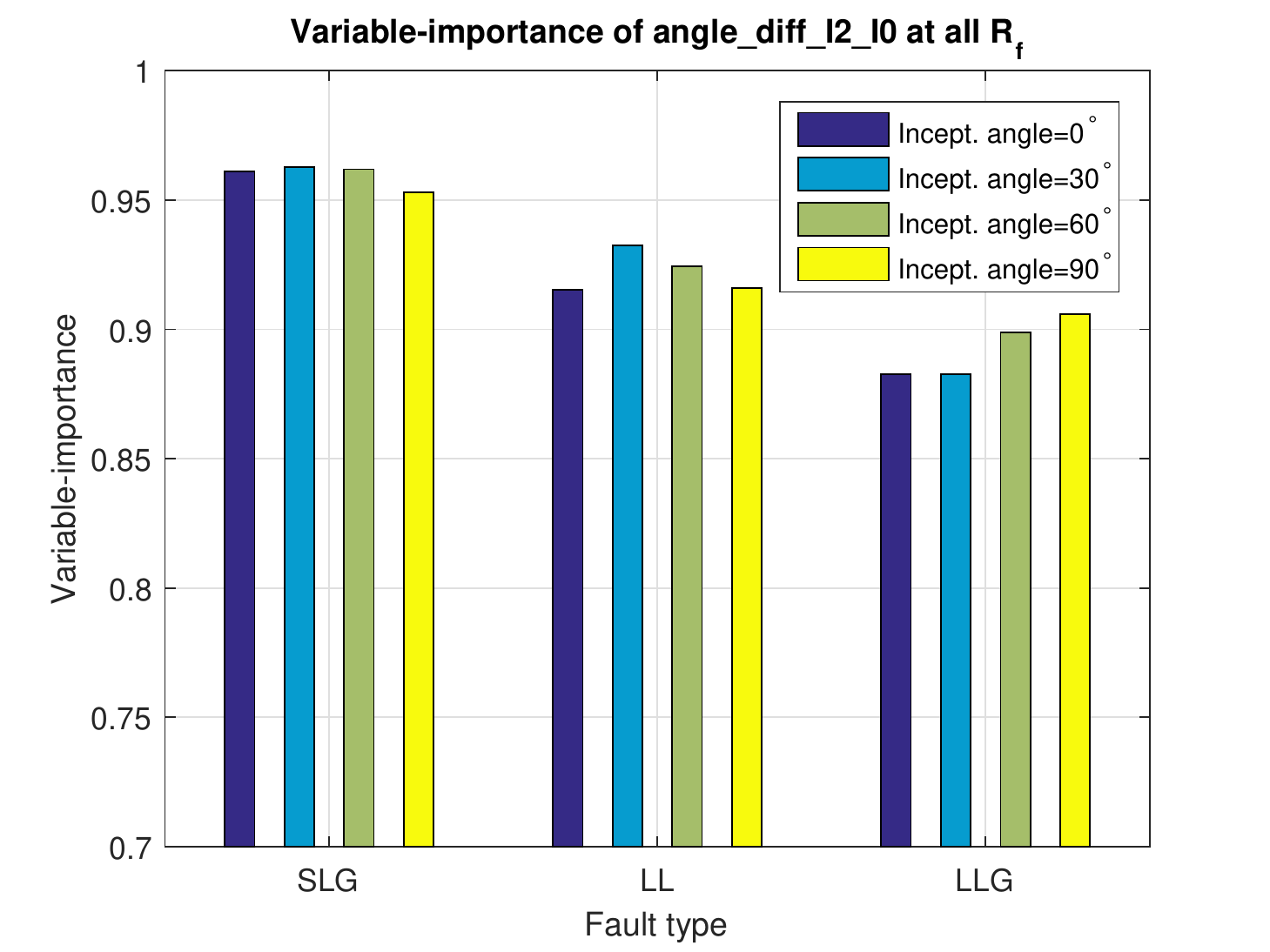}} 
     \subfloat[Feature $KF\_V\_cos\_H3$ at different fault inception angles.]{\label{fig:inceptionangle_KF_V_cos_H3}\includegraphics[width=3.0in]{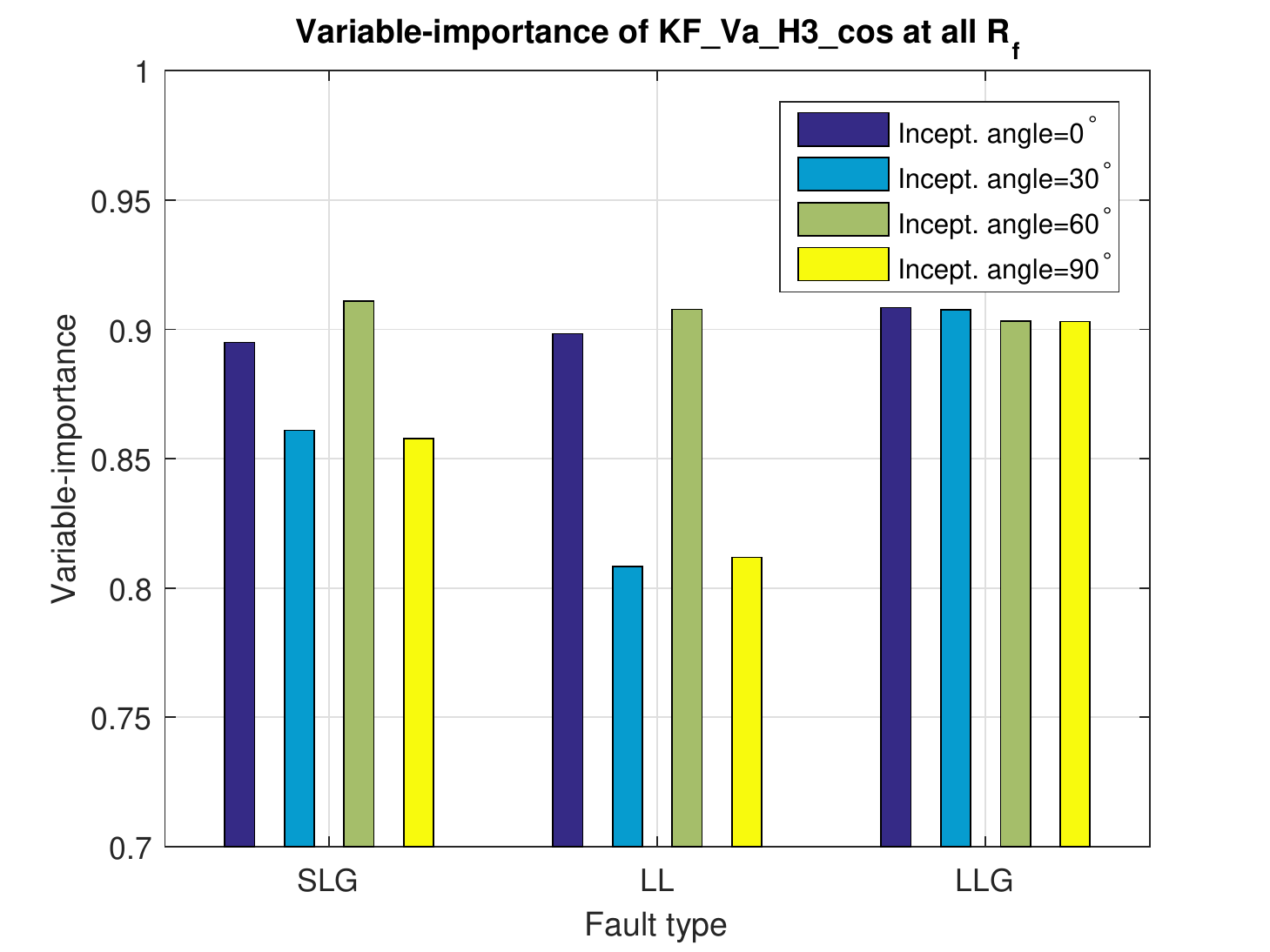}} \\
     \subfloat[LLLG fault features at different fault inception angles.]{\label{fig:inceptionangle_LLLG}\includegraphics[width=3.0in]{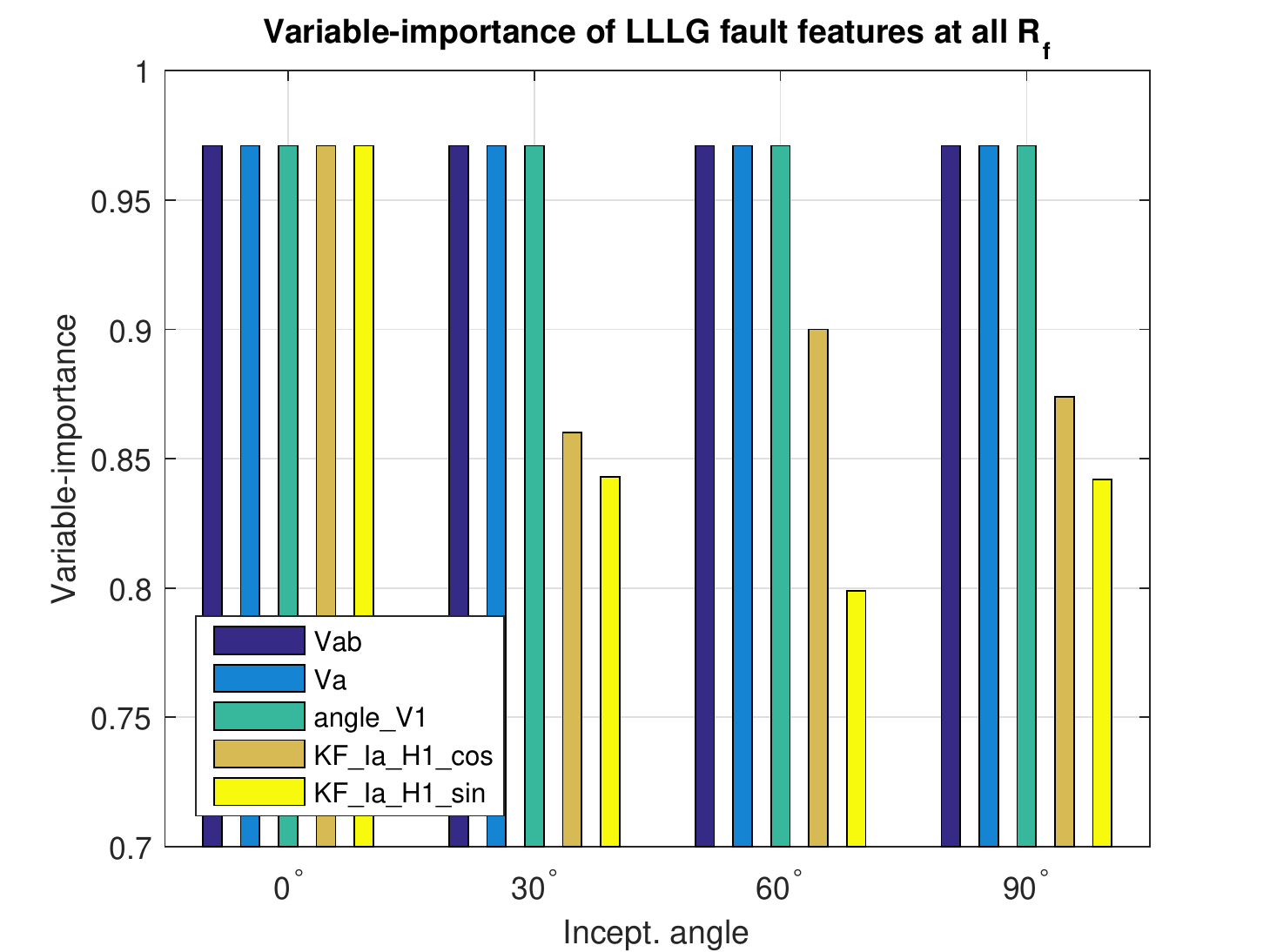}} 
     \subfloat[All $R_f$ at different fault locations.]{\label{fig:location_allRf}\includegraphics[width=3.0in]{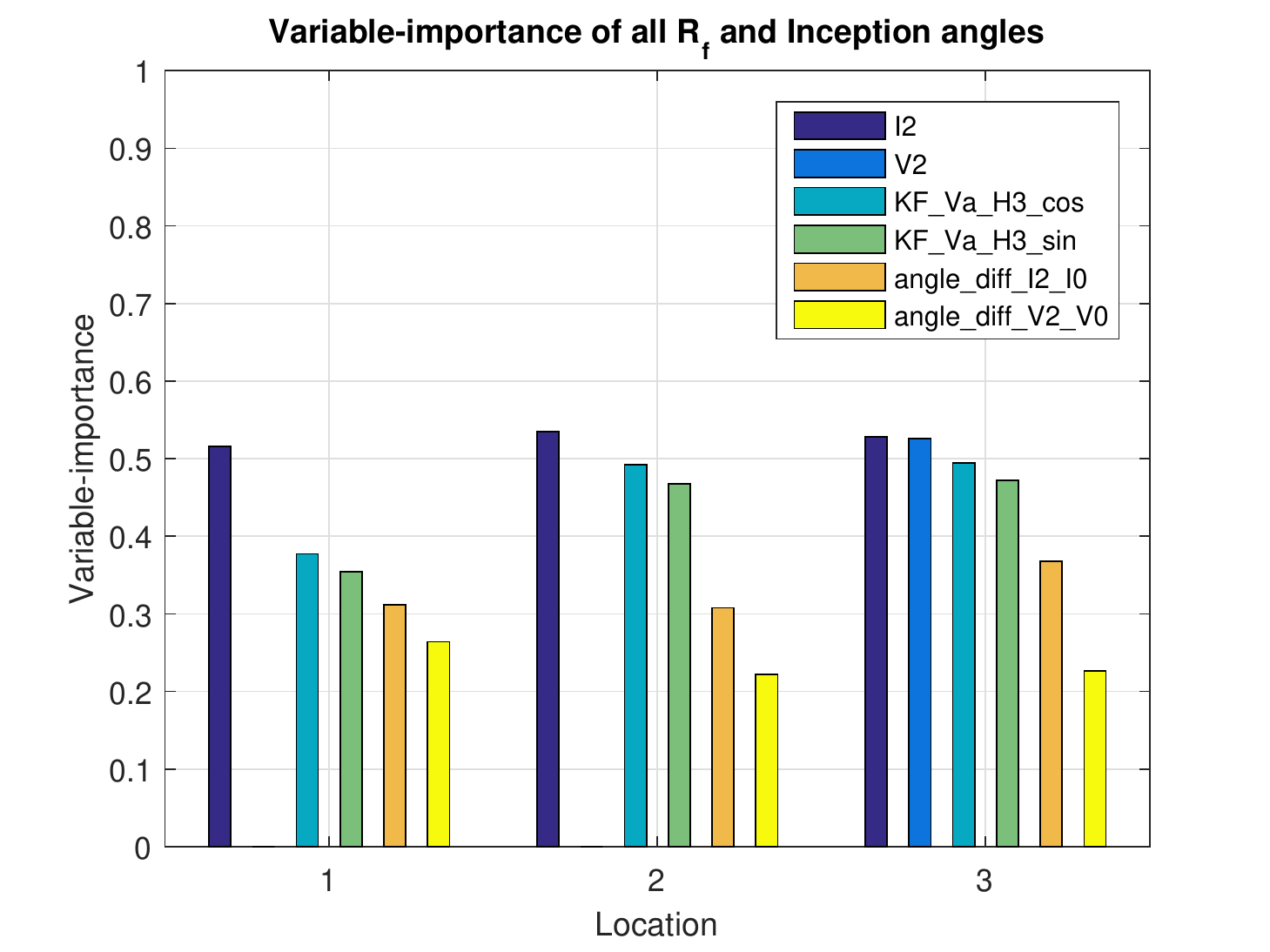}}
       \caption{Variable-importance at different fault inception angles and locations.} 
     \label{fig:variableimportance}
     \vspace{-4mm}
 \end{figure*}

The affect of fault inception angle is examined as well in this study. The results of unbalanced faults and three phase faults are selectively shown in Fig. \ref{fig:variableimportance}(a)-(c) respectively. The results in these figures include a varying impedance from $30$ $\Omega$ to $500$ $\Omega$. 





The fault inception angle is an insignificant factor that can perturb variable importance. The angles of $30$\textdegree \space and $60$\textdegree \space result in a subtle decrease in the variable of importance of the KF estimated third harmonic, but the change is limited. For an LLLG fault, the first order harmonic components of current estimated by KF have a performance drop in non-zero angles.

%

%

\subsubsection{Fault Location}
\label{sec:faultlocationHIFtest}
The variable-importance of the features in EFS is presented at three fault locations (bus numbers refer to Appendix \ref{sec:appbenchmarksys}): 

\begin{itemize}
\item Location $1$: Fault near Bus B-$3$; 
\item Location $2$: Fault near Bus B-$11$; 
\item Location $3$: Fault near Bus B-$19$; 
\end{itemize}

The result is demonstrated in Fig. \ref{fig:variableimportance}(d), including all fault impedance and all fault inception angles in Table \ref{tab:eventcategory}. The feature of negative sequence current keeps being unaffected at each location. However, the negative sequence of voltage is so low at location $1$ and $2$ that the variable of importance becomes almost zero. As the strong voltage source from the substation is ideally balanced, the negative sequence voltage deviation contributed from the HIF is weak. Location $3$ is far from the substation, so the negative sequence voltage becomes a good HIF indicator again. To a negligible extent, it is similar for the variable of importance performance of other features: the further the fault is, the less compromised the features are.

\vspace{-3mm}

\subsection{Testing Results of the Effective Feature Set (EFS)}

The proposed EFS is tested under the aforementioned conditions in Table \ref{tab:eventcategory}, but with data set on different events. $1944$ HIF events and another $1944$ non-HIF events (they are unnecessary to be the same number) are simulated for the training of the HIF detector. The types of testing events are similar to those of training events, but at different locations or with different parameter values. There are totally $972$ HIF and $972$ non-HIF events in the testing. In addition, five classical classifiers (Naive Bayes, Support Vector Machine, k-nearest neighbour, J48 and random forest) from Weka are compared in order to find the best classifier.

The results with proposed EFS are shown in Table \ref{tab:classifiers77}, where the accuracy is the ratio of correctly classified event number to the total number of events, dependability index (DI) is the ratio of the detected HIF events to the total number of HIF events and security index (SI) is the ratio of the of detected non-HIF events to the total number of non-HIF events. To limit problems such as over-fitting and inaccuracy in prediction, each classifier model is acquired through the $10$-folder cross-validation.

\begin{table}[!hbt]
	\renewcommand{\arraystretch}{1.3}

	\caption{Performance of HIF detection with the EFS.}
	\label{tab:classifiers77}
	\centering
	\begin{tabular}{cccc}
		\hline
		\hline
		Classifier & Accuracy (\%) & DI (\%)& SI (\%)\\
		\hline
		NaiveBayes & 78.0 & 73.0 & 82.9\\
		\hline
		LibSVM (Gaussian kernel) & 91.9 & 89.6 & 94.1 \\
		\hline
		k-nearest neighbour & 98.0 & 97.7 & 98.3 \\
		\hline
		J48 & 99.4 & 99.5 & 99.3 \\
		\hline
		RandomForest & 99.7 & 99.7 & 99.8 \\
		\hline
		\hline
	\end{tabular}
	\end{table}

The EFS is tested in the benchmark system under two typical $X/R$ ratios: $10$ and $22.5$. Both scenarios result in the same EFS. The $X/R$ ratio is not an important factor that affects the proposed method since it does not affect the system unbalance level.

\section{Conclusions}
\label{sec:concludeHIF}

This paper proposes a new framework for HIF detection and classification. By introducing the MDL-based algorithm to rank a pool of systematically designed features, an effective feature set is generated. The detection capability of such a ranked feature set is evaluated through a comprehensive fault analysis on different scenarios. Furthermore, an applicable logic is recommended based on the extensively used techniques of DFT and KF as well as easily implementable logic gates. It is shown that the proposed method achieves significantly enhanced performance in HIF detection with the effective feature set and tree-based classifier such as the random forest.  


\ifCLASSOPTIONcaptionsoff
  \newpage
\fi


\appendices

\section{HIF model}
\label{sec:Journal2017AppHIFmodel}

\begin{figure}[h]
	\centering
	\includegraphics[width=1.6in]{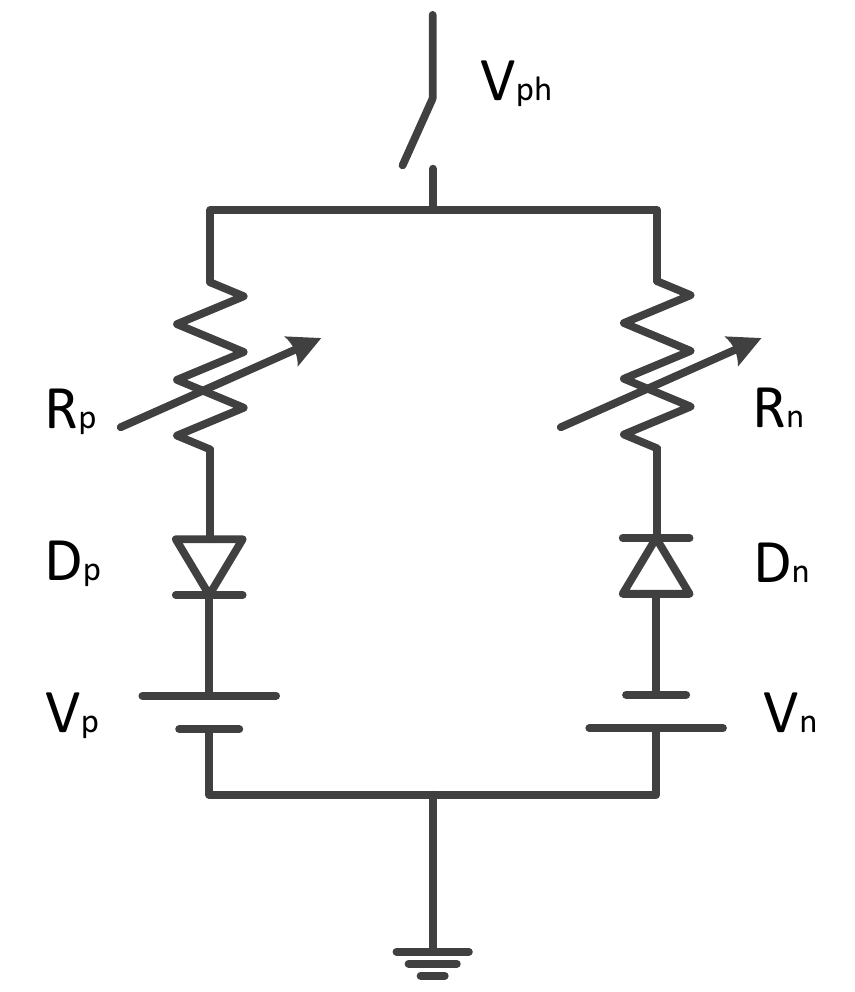}
	\caption{HIF two anti-parallel dc-source model.}
	\label{fig:hifantidcmodel}
	\vspace{-5mm}
\end{figure}

Fig. \ref{fig:hifantidcmodel} shows the HIF model used in this paper. This model connects one phase of the power line to the ground. Two variable resistors are both changing randomly and model the dynamic arcing resistance. Two sets of diodes and DC sources are connected in an anti-parallel configuration. The two DC sources are randomly varying as well, which model the asymmetric nature of HIF. The positive half cycle of HIF current is achieved when $V_{ph}>V_p$, while negative half cycle when $V_{ph}<V_n$. When $V_n<V_{ph}<V_p$, the current equals to zero, which represents the period of arc extinction. In order to generate a fault current between $10$ and $100$ A in the benchmark system, we adopt the model settings in Table \ref{tab:HIFmodelSetting}. 
 
\begin{table}[hbt]
	\renewcommand{\arraystretch}{1.3}
	\caption{HIF model settings.}
	\label{tab:HIFmodelSetting}
	\centering
	\begin{tabular}{ccc}
		\hline		
		\hline
		Component & Value range & Values change every\\
		\hline
		$V_p$ & $5 \sim 6$ kV & 0.1 ms\\
		\hline
		$V_n$ & $7 \sim 8$ kV & 0.1 ms\\
		\hline
		$R_p$ & $200 \sim 1500$ $\Omega$ & 0.1 ms\\
		\hline
		$R_n$ & $200 \sim 1500$ $\Omega$ & 0.1 ms\\
		\hline
		\hline		
	\end{tabular}
	\vspace{-5mm}
\end{table}

\section{Derivation of the Minimum Description Length}
\label{sec:variableimportanceappendix}

Assuming that all features are discrete, the objective is to find the best features that maximize the selection measure. Let C, A and V denote the number of classes, the number of features, and the number of values of the given feature. With this notation, we show in the following the entropy of the classes ($H_C$), the values of the given feature ($H_A$), the joint events class-feature value ($H_{CA}$), and the classes given the value of the attribute ($H_{C|A}$).

\begin{align}
H_C&=-\sum_{i} p_{i.} \log p_{i.}, &H_A=-\sum_{j} p_{.j} \log p_{.j},\nonumber\\
H_{CA}&=-\sum_{i} \sum_{j} p_{ij} \log p_{ij}, &H_{C|A}=H_{CA}-H_A,\nonumber
\end{align}
\vspace{-5mm}

where $p_{ij}=n_{ij}/n_{..}$, $p_{i.}=n_{i.}/n_{..}$, $p_{.j}=n_{.j}/n_{..}$ and $p_{i|j}=n_{ij}/n_{.j}$. ``$n..$" denotes the number of training instances and ``$n_{i.}$" is the number of training instances from class $C_i$, $n_{.j}$ is the number of instances with the $j$-th value of the given attribute, and $n_{ij}$ is the number of instances from class $C_i$ and with the $j$-th value of the given attribute.

The approximation of the total number of bits that are needed to encode the classes of $n_{..}$ is:
\begin{equation}
\text{Prior MDL}'=n..H_C+\log \tbinom{n..+C-1}{C-1},
\end{equation}
and the approximation of the number of bits to encode the classes of examples in all subsets corresponding to all values of the selected attribute is:
\begin{equation}
\text{Post MDL}'=\sum_{j} n_{.j}H_{C|j}+\sum_{j} \log \tbinom{n_{.j}+C-1}{C-1}+\log A.\nonumber
\end{equation}

The last term ($\log A$) is needed to encode the selection of an attribute among $A$ attributes. However, this term is constant for a given selection problem and can be ignored. The first term equals $n_{..}H_{C|A}$. Therefore, the MDL$'$ measure evaluates the average compression (per instance) of the message by an attribute. The measure is defined by the following difference, Prior MDL$'$ $-$ Post MDL$'$, normalized with $n_{..}$:
\begin{align}\label{eq:paretomle1}
\text{MDL}'=\text{Gain}+\frac{1}{n..} \Big( \log \tbinom{n..+C-1}{C-1} \\ -\sum_{j} \log \tbinom{n_{.j}+C-1}{C-1}\Big).
\end{align}

However, entropy $H_C$ can be used to derive MDL$'$ if the messages are of arbitrary length. If the length of the message is known, the more optimal coding uses the
logarithm of all possible combinations of class labels for given probability distribution:

\vspace{-4mm}
\begin{equation}
\text{Prior MDL}=\tbinom{n..}{n_{1.},...,n_{C.}}+\log \tbinom{n..+C-1}{C-1}
\end{equation}
\vspace{-4mm}

Similarly, if we use the priori minus the posterior of the $MDL$, equation (\ref{eq:paretomle2}) is obtained.

\section{Benchmark System}
\label{sec:appbenchmarksys}

\begin{figure*}[!htb]
	\centering
	\includegraphics[width=6.5in]{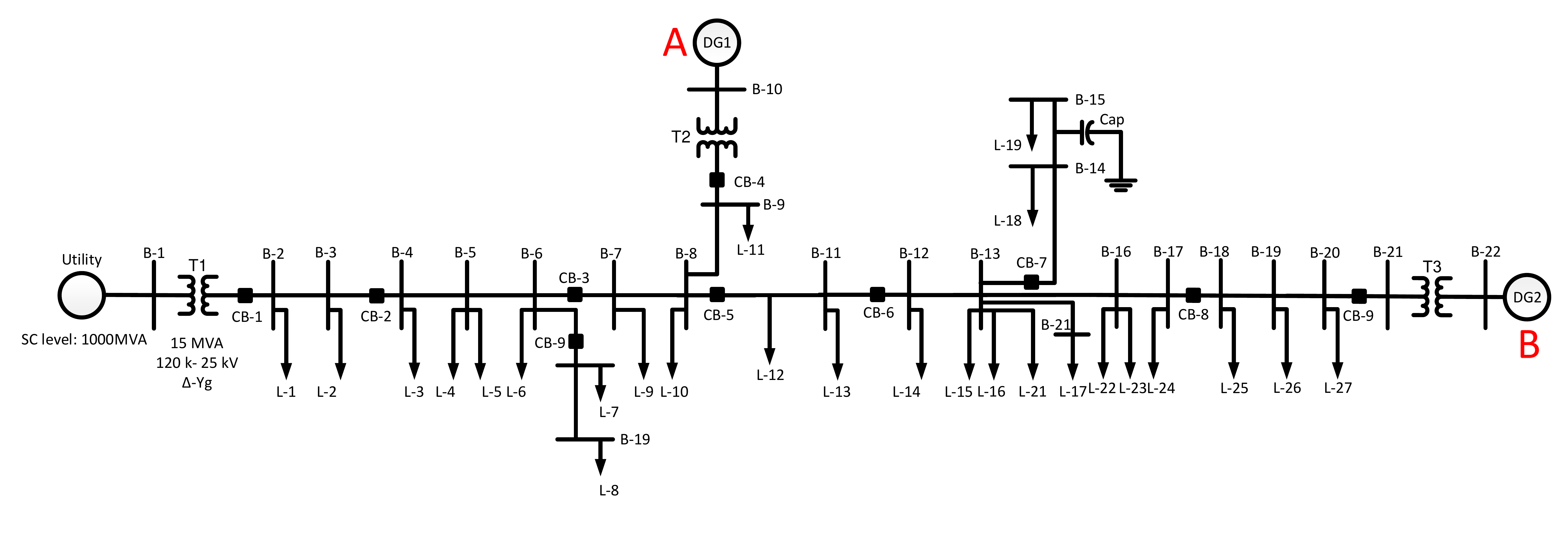}
	\centering
	\caption{Single line diagram of distribution feeder under study.}
	\label{fig:feeder_HIF}
\end{figure*}

The benchmark system is presented in Fig. \ref{fig:feeder_HIF}, the details of which can be found in \cite{ref:DavyZhuang}.

%


%


\bibliographystyle{IEEEtran}
\bibliography{IEEEabrv,HIF_onlineSVM}

%
%
\begin{IEEEbiography}[{\includegraphics[width=1.2in,height=1.35in,clip,keepaspectratio]{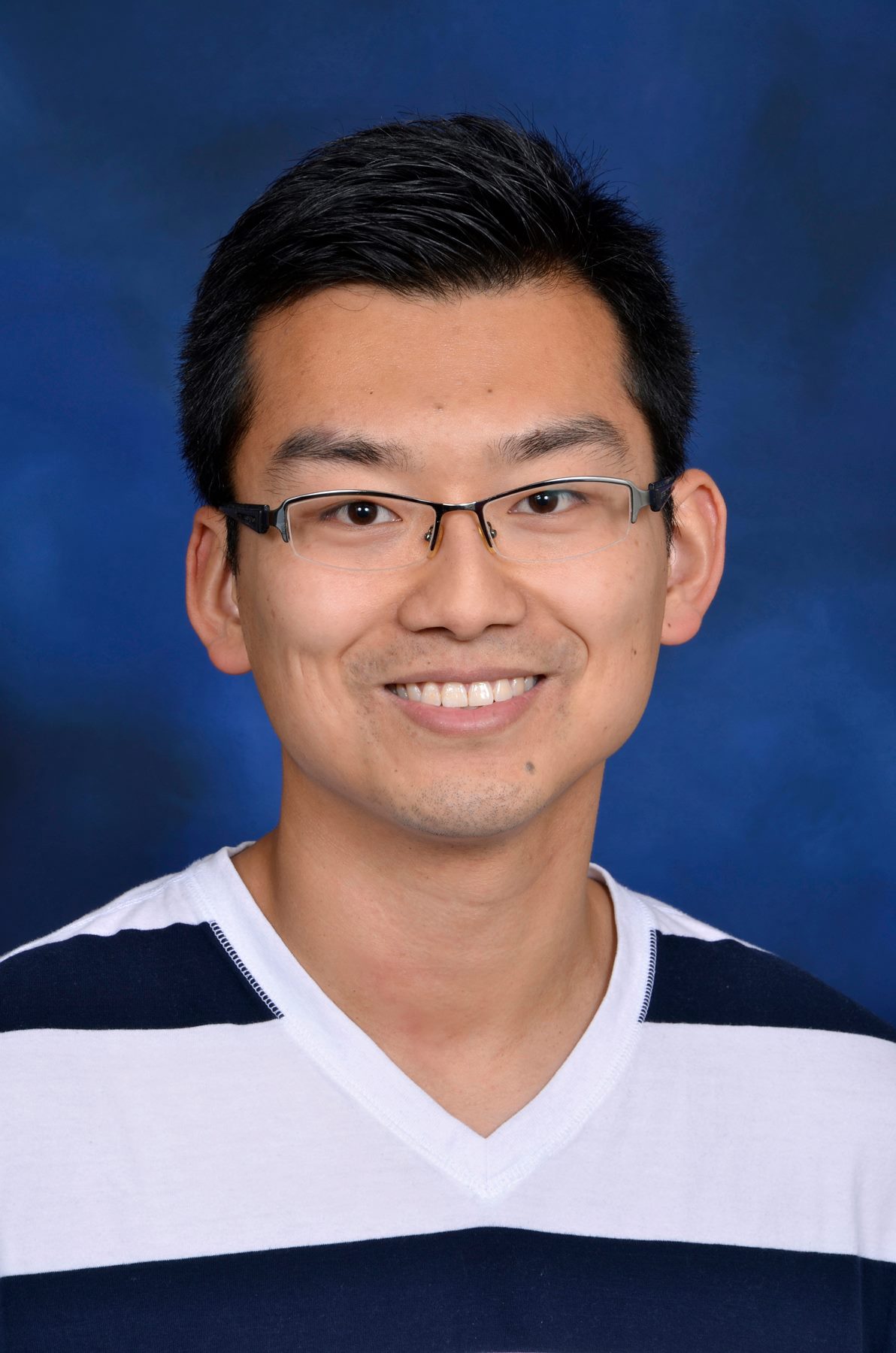}}]{Qiushi Cui} (S'10-M'18) received the M.Sc. degree from Illinois Institute of Technology in 2012, and the Ph.D. degree from McGill University in 2018, both in Electric Engineering. He was a research engineer at OPAL-RT Technologies Inc. since Nov. 2015, and sponsored by Canada MITACS Accelerate Research Program at the same company until Nov. 2017. Since Jan. 2018, he joined Arizona State University as a postdoctoral researcher.  

His research interests are in the areas of big data applications in power system protection, power system modeling, microgrid, EV charging station placement, and real-time simulation in power systems. 

\end{IEEEbiography}

\begin{IEEEbiography}[{\includegraphics[width=1in,height=1.25in,clip,keepaspectratio]{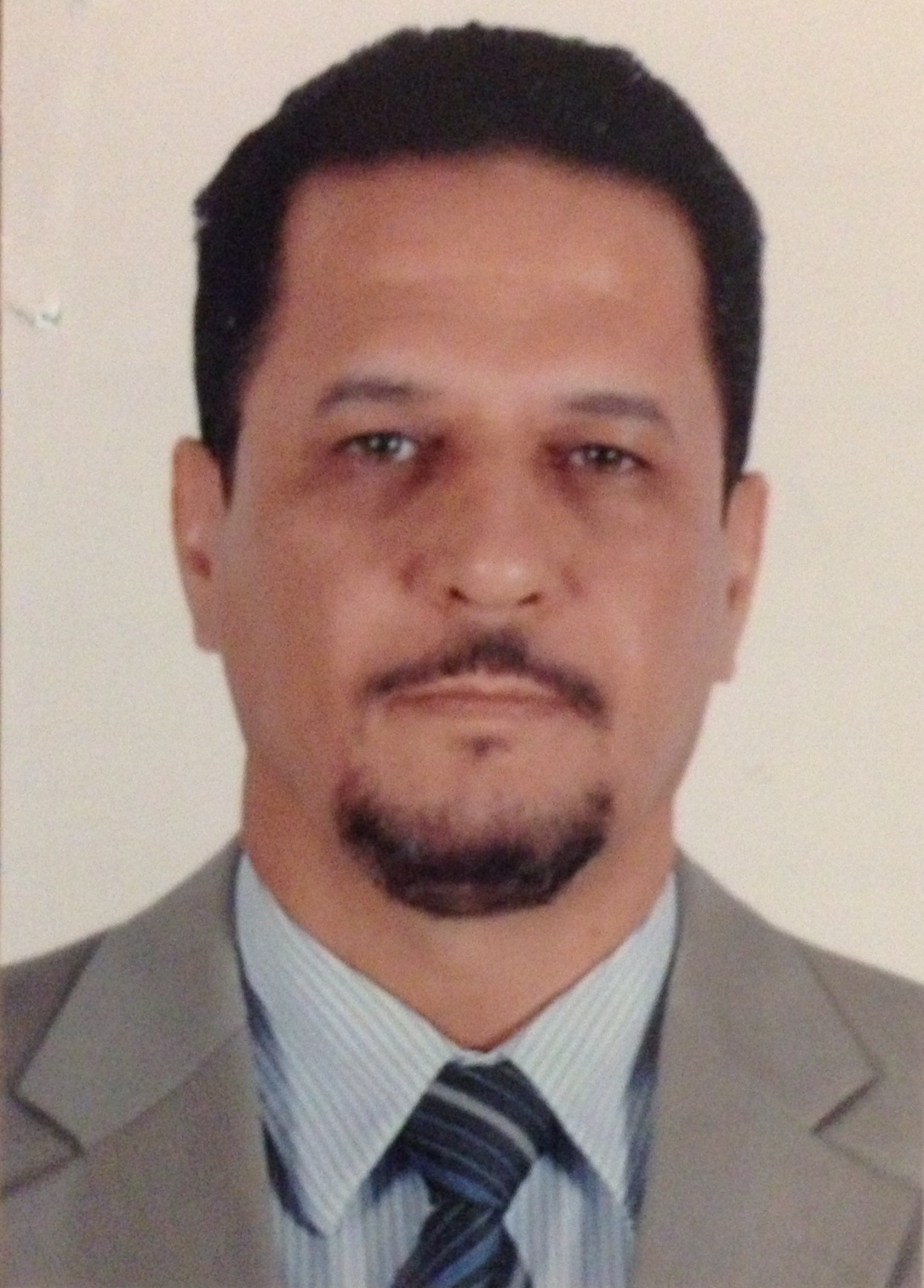}}]{Khalil El-Arroudi} (S'99-M'06) received the B.Sc and M.Sc. degrees from Garyounis University, Libya, and the Ph.D. degree from McGill University, Canada, in 1986, 1994, and 2004 respectively. He has twenty-three years of experience in power system protection, and system operation. At the General Electricity Company of Libya (GECOL), he has occupied different managerial positions including the General Manager of System Operation, and the Chairman and Managing Director.

He has been a member and chair in different international Mediterranean interconnection committees including the LTAM-ENTSOE interconnection (Maghrebian-European projects), the Arab Union of Electricity, the Mediterranean Electricity Network (MEDELEC), and the OME (Observatoire M\'editerran\'een de l\'Energie). He holds two USA Patents US8200372B2 related to transmission systems and distributed generations. He is presently a research staff at McGill University. His research interests include power system protection, distributed generation applications, microgrids, smart grids, and applications of intelligent applications in power systems.

\end{IEEEbiography}

\begin{IEEEbiography}[{\includegraphics[width=1in,height=1.25in,clip,keepaspectratio]{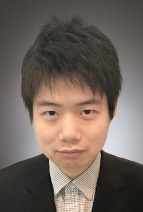}}]{Yang Weng} (M'14) received the B.E. degree in electrical engineering from Huazhong University of Science and Technology, Wuhan, China; the M.Sc. degree in statistics from the University of Illinois at Chicago, Chicago, IL, USA; and the M.Sc. degree in machine learning of computer science and M.E. and Ph.D. degrees in electrical and computer engineering from Carnegie Mellon University (CMU), Pittsburgh, PA, USA.

After finishing his Ph.D., he joined Stanford University, Stanford, CA, USA, as the TomKat Fellow for Sustainable Energy. He is currently an Assistant Professor of electrical, computer and energy engineering at Arizona State University (ASU), Tempe, AZ, USA. His research interest is in the interdisciplinary area of power systems, machine learning, and renewable integration.

Dr. Weng received the CMU Dean’s Graduate Fellowship in 2010, the Best Paper Award at the International Conference on Smart Grid Communication (SGC) in 2012, the first ranking paper of SGC in 2013, Best Papers at the Power and Energy Society General Meeting in 2014, ABB fellowship in 2014, and Golden Best Paper Award at the International Conference on Probabilistic Methods Applied to Power Systems in 2016.

\end{IEEEbiography}

\end{document}